%% file: main.tex
\documentclass[letterpaper,twocolumn,10pt]{article}
\usepackage{usenix}

\usepackage{tikz}
\usepackage{amsmath}
\usepackage{lipsum}
\usepackage{multirow}
\usepackage{threeparttable}
\usepackage[table]{xcolor}
\usepackage{pdflscape}

\usepackage{hyperref}
\usepackage{xurl}

\usepackage{xstring}
\usepackage{booktabs}
\usepackage[most]{tcolorbox}
\usepackage{arydshln}
\usepackage{listings}
\usepackage{amsfonts}
\definecolor[named]{ACMBlue}{cmyk}{1,0.1,0,0.1}
\definecolor[named]{ACMOrange}{cmyk}{0,0.42,1,0.01}
\definecolor[named]{ACMRed}{cmyk}{0,0.90,0.86,0}
\definecolor[named]{ACMGreen}{cmyk}{0.20,0,1,0.19}
\definecolor[named]{ACMPurple}{cmyk}{0.55,1,0,0.15}

\newcommand{\truncate}[1]{%
  \StrLeft{#1}{8}...\StrRight{#1}{8}
}

\newcommand{\AddrHrefEthereum}[2][blue]{\href{https://etherscan.io/address/#2}{\color{#1}{\truncate{#2}}}}%

\newcommand{\featheader}[1]{\emph{\textbf{{#1}.}}}

\newtcolorbox{findingbox}[1][]{%
  colback=gray!5,      % Light blue background
  colframe=black,    % Darker blue frame
  boxrule=1pt,       % Frame thickness
  arc=0pt,             % Rounded corners
  auto outer arc,
  left=2mm, right=2mm,
  top=1mm, bottom=1mm,
  fontupper=\footnotesize,
  #1
}

\begin{document}
%-------------------------------------------------------------------------------

%don't want date printed
\date{}

\title{\Large \bf Exploiting Liquidity Exhaustion Attacks in Intent-Based  Cross-Chain Bridges}

%\author{Anonymous Submission}
%for single author (just remove % characters)
\author{
{\rm André Augusto}\\
INESC-ID, Instituto Superior Técnico, \\University of Lisbon
\and
{\rm Christof Ferreira Torres}\\
INESC-ID, Instituto Superior Técnico, \\University of Lisbon
\and
{\rm André Vasconcelos}\\
INESC-ID, Instituto Superior Técnico, \\University of Lisbon
\and
{\rm Miguel Correia}\\
INESC-ID, Instituto Superior Técnico, \\University of Lisbon
} % end author

% copy the following lines to add more authors
% \and
% {\rm Name}\\
%Name Institution

\maketitle
%\tableofcontents

\begin{abstract}

Intent-based cross-chain bridges have emerged as an alternative to traditional interoperability protocols by allowing off-chain entities (\emph{solvers}) to immediately fulfill users' orders by fronting their own liquidity. While improving user experience, this approach introduces new systemic risks, such as solver liquidity concentration and delayed settlement. %: before solver and protocol availability depend on the liquidity of a small set of solvers.
In this paper, we propose a new class of attacks called \emph{liquidity exhaustion attacks} and a replay-based parameterized attack simulation framework. We analyze 3.5 million cross-chain intents that moved \$9.24B worth of tokens between June and November 2025 across three major protocols (Mayan Swift, Across, and deBridge), spanning nine blockchains.

For rational attackers, our results show that protocols with higher solver profitability, such as deBridge, are vulnerable under current parameters: 210 historical attack instances yield a mean net profit of \$286.14, with 80.5\% of attacks profitable. In contrast, Across remains robust in all tested configurations due to low solver margins and very high liquidity, while Mayan Swift is generally secure but becomes vulnerable under stress-test conditions. Under byzantine attacks, we show that it is possible to suppress availability across all protocols, causing dozens of failed intents and solver profit losses of up to \$978 roughly every 16 minutes. Finally, we propose an optimized attack strategy that exploits patterns in the data to reduce attack costs by up to 90.5\% compared to the baseline, lowering the barrier to liquidity exhaustion attacks.% only through the exploitation of solver participation patterns.

\end{abstract}

\input{paper-body}

\begin{comment}
\section*{Acknowledgments}

\end{comment}

% optional clearing of the page
%\cleardoublepage
\appendix
%\section*{Ethical Considerations}
%All experiments rely exclusively on historical, publicly available on-chain data and were conducted offline. We did not deploy smart contracts, submit transactions, manipulate protocol behavior, or interfere with live systems, ensuring that our research did not cause any impact. The analysis uses only pseudonymous blockchain data and does not involve personally identifiable information. Our objective is to strengthen the ecosystem resilience by identifying systemic risks before they are exploited in practice. We have communicated our findings to the teams of each protocol studied so that they can review and acknowledge the results.

% optional clearing of the page
%\cleardoublepage

%\section*{Open Science}
%To support openness and verifiability, the dataset and code used in this paper will be released publicly after the double-blind review.

% optional clearing of the page
%\cleardoublepage
\bibliographystyle{plainurl}
\bibliography{references}

\section{Cross-Chain Transactions Data Format}\label{appendix: data_model}
Each cross-chain transaction contains information about the intent issued by the user, the fulfillment by the solver, and the repayment transaction. Table~\ref{tab:data-model} shows a simplified description of the cross-chain transactions data structure.
\input{tables/data-model}

\section{Solver Addresses}\label{appendix: addresses}
Table~\ref{tab: solver_addreses} presents all solver addresses identified in our dataset, that fulfilled user intents in Ethereum.
\input{tables/solver-addresses}

%%%%%%%%%%%%%%%%%%%%%%%%%%%%%%%%%%%%%%%%%%%%%%%%%%%%%%%%%%%%%%%%%%%%%%%%%%%%%%%%
\end{document}

%% file: paper-body.tex
\section{Introduction}
\label{sec: introduction}

\begin{figure}
    \centering
    \includegraphics[width=\columnwidth]{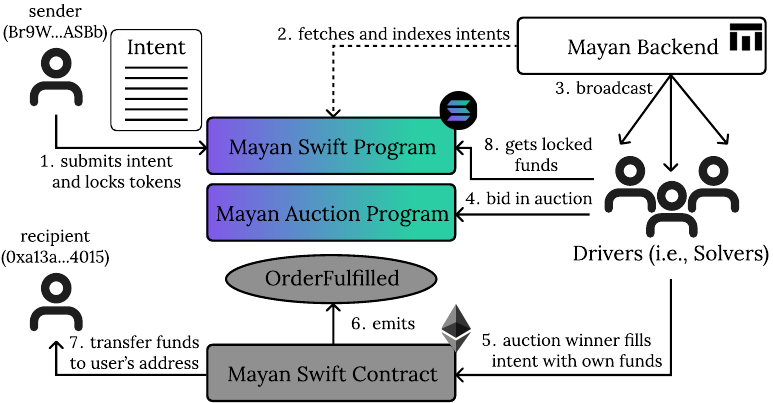}
    \caption{Cross-chain transaction flow from Solana to Ethereum using Mayan Swift's bridge (a Wormhole bridge).}
    \label{fig: mayan-bridge}
\end{figure}

% Intents having an increased dominance
The decentralized finance (DeFi) ecosystem has recently undergone a significant architectural shift with the rise of intent-based protocols. Rather than specifying explicit transaction steps, users now specify the desired end state (i.e., their \textit{intent}), which is executed by specialized off-chain actors called \emph{solvers} that compete to fulfill these orders in an auction~\cite{lifi_intents}. This paradigm, initially adopted in decentralized exchanges (DEXes), is rapidly expanding to cross-chain systems, where its complexity and security implications become more pronounced.

% Is this too descriptive for the introduction, and should be moved to the background?
In cross-chain environments, intent-based designs promise a faster movement of assets across heterogeneous blockchains. Instead of dealing with the traditional ``lock–relay–unlock'' pipeline (or any other variation such as lock-mint, or burn-mint), users submit intents such as: ``swap x units of token X on chain A for y units of token Y on chain B''~\cite{openintents.xyz}. Solvers bid in auctions to fulfill these intents as fast as possible by fronting their own liquidityA on the destination chain. Later on, solvers reclaim repayment through an intent settlement mechanism that verifies the correctness of the fill in relation to the user's intent. High-profile protocols such as Across~\cite{across} (used by Uniswap~\cite{uniswap_across}), Mayan Swift (the second largest Wormhole-based bridge behind Portal, with the highest volume between Solana and Ethereum~\cite{wormhole_protocol_stats}), and deBridge~\cite{debridge} have adopted this design due to its low latency and improved user experience.

% Inversion of execution order compared to vanilla-bridges. Description of the new security risks
There is an inversion of the execution order of intent-based bridges that marks a fundamental departure from classical bridges. Traditional bridges move the user's funds first and release assets on the destination chain only after finality is achieved at the source and whenever a set of signatures or proofs is computed. Intent-based systems reverse this sequence: solvers deliver assets immediately, while verification occurs afterward (cf. Figure~\ref{fig: mayan-bridge}). The protocol auction specifies the winning solver that must fulfill the intent. This design improves latency and user experience, but introduces new systemic risks. In particular, solvers must maintain substantial locked liquidity, and settlement delays or failures shift the risk from the user to the solver. As a result, the safety and liveness of the protocol are critically dependent on the total liquidity available and the availability of these solvers.

% Where this paper fits. The gap
This paper aims to raise awareness for a new class of attacks and explores an under-researched topic: the susceptibility of intent-based cross-chain protocols, or specific solvers, to suffer \emph{liquidity exhaustion attacks}. Since solvers front capital before being reimbursed, their liquidity can be temporarily drained, either accidentally, strategically by rational adversaries exploiting profit conditions, or intentionally by byzantine actors aiming to disrupt service. To our knowledge, this is the first paper that investigates not only such attacks but also cross-chain intents in general.
%\mpc{alterei a frase acima pois acho que é isso (dizia apenas "To our knowledge, this is the first paper that investigates cross-chain intents.")}
%\andre{concordo}

We summarize the contributions as follows:

\begin{itemize}\setlength\itemsep{-0.1em}
    \item \textbf{First empirical measurement} of solver liquidity, settlement delays, and real-world intent flows. We collect data from Jun 1, 2025 to Nov 1, 2025, totaling 3.5 million intents moving more than \$9.24 billion in three major intent-based bridges across nine blockchains.
    \item \textbf{Analyzing vulnerabilities} that arise when adversaries submit sequences of intents designed to temporarily decrease solver liquidity, preventing honest solvers from participating until the intent refund.
    \item \textbf{Evaluating attacks under both rational and byzantine models}, showing how liquidity exhaustion can halt or severely degrade service depending on protocol parameters, settlement latencies, and solver heterogeneity. %Additionally, we show that major protocols have their \textbf{security dependent on external usage patterns, instead of internal parameters}.
    \item \textbf{Proposing defense guidelines and mitigation strategies} for protocol operators and solvers to minimize the risks associated with such attacks.
\end{itemize}

This paper is structured as follows. Section~\ref{sec: background} presents an overview of cross-chain bridges and a theoretical comparison of each protocol analyzed in the paper. In Section~\ref{sec: liq_exhaustion_attacks} we propose the \emph{Liquidity Exhaustion Attacks}. Section~\ref{sec: data_collection_and_analysis} overviews our data collection process and identifies some key vulnerabilities that make such attacks possible. Section~\ref{sec:simulation_framework} describes our simulation framework used in Sections~\ref{sec: baseline_strategy},~\ref{sec: byzantine_adversary}, and~\ref{sec: targeted_attack} assuming both rational and byzantine adversaries. Sections~\ref{sec: defense_strategies} and~\ref{sec:discussion} contain some defense strategies and the discussion of our findings. Finally, Sections~\ref{sec: related_work} and~\ref{sec: conclusion} outline the related work and the conclusion of the paper.

%\lipsum[1-2]

%https://x.com/ArrakisFinance/status/1864038151150596125

%See Arjund emails...

%This is becoming increasingly important in the ecossystem, as more and more organizations are going cross-chain, including Uniswap with the new "Unichain", and as we can see by the increase in the total value locked in these protocols.
%see https://www.linkedin.com/pulse/7-things-you-need-know-unichain-uniswaps-latest-l2-krzysztof-gogol-hnmae/

% uniswap is using Across - https://blog.uniswap.org/permissionless-transferring-is-now-live

\section{Cross-Chain Bridges and Intents}\label{sec: background}

Cross-chain bridges enable interoperability between heterogeneous blockchains~\cite{belchior2021survey}. Because blockchains maintain independent consensus, execution, and data availability layers, assets on one chain cannot be transferred directly to another. Bridges address this limitation by providing a mechanism to move value or messages across blockchains, typically through \emph{lock-mint}, \emph{burn-mint}, or \emph{lock-unlock} schemes~\cite{belchior2023you}. In these designs, a user's asset on the source chain is locked or destroyed, and a corresponding representation is released or minted on the destination chain once verification has been performed, often in the form of validator signatures, light-client proofs, or threshold attestations. Although widely deployed, traditional bridges inherently suffer from high latency~\cite{augusto2025xchaindatagen}.

\subsection{Cross-Chain Intents}
Intent-based interoperability introduces a fundamentally different model for cross-chain execution. Instead of specifying concrete transaction steps, users express their desired final state (an \emph{intent}) and delegate the execution to a market of competing off-chain actors. This paradigm shifts operational complexity away from users and towards solvers, who are capable of satisfying the intent across blockchains. Figure~\ref{fig: mayan-bridge} depicts the flow of information in Mayan Swift, a Wormhole~\cite{wormhole}-based bridge known for bridging EVM-based blockchains and Solana.

The terminology used in the paper is defined below:

\begin{itemize}\setlength\itemsep{-0.1em}
    \item \textbf{Intents.} A high-level user specification of a desired final state across chains, without prescribing the concrete steps required to achieve it.
    
    \item \textbf{Solvers.} Off-chain actors that compete to fulfill intents by fronting liquidity and executing the cross-chain operations.
    
    \item \textbf{Intent Fulfillment.} The process in which a solver delivers the requested assets to the user on the destination chain.
    
    \item \textbf{Intent Settlement.} Verification that the solver's fulfillment (sometimes mentioned as just \emph{fill}) matches the user's intent and triggers the refund process.
%\mpc{acima o termo fill é esquisito ou está pouco claro}

    \item \textbf{Solver Refunds.} The asynchronous repayments solvers receive after settlement, releasing the liquidity they fronted during fulfillment.
\end{itemize}

\subsection{Solver Behavior and Liquidity}
Table~\ref{tab: protocols-summary} summarizes the key differences among the three protocols analyzed in this paper, including usage patterns, solver behavior, fulfillment mechanisms, and settlement delays. The table is derived from our analysis of the collected dataset (see Section~\ref{sec: data_collection_and_analysis} for details).

\input{tables/protocols-summary}

\featheader{Solver Selection} Protocols vary in their solver selection mechanisms and solver concentration. All protocols allow open participation in the auctions. However, Mayan Swift employs an English auction with a relatively distributed solver set, while Across and deBridge follow a first-come-first-served (FCFS) model. deBridge exhibits strong solver centralization, with a single solver responsible for 94\% of fulfillments. Across has a bigger set of solvers and the top solver is responsible for 24\% of the fulfillments.%Such concentration directly affects how liquidity is deployed and reused across intents.

\featheader{Solver Refund Times} Median refund times range from approximately 1{,}000 seconds (in Mayan and deBridge) to more than two hours (in Across), during which the solver capital remains locked. The massive difference between Across and the other protocols is due to its reliance on an optimistic oracle with 2h fraud proof window. %Longer refund times reduce effective available liquidity, even when total solver liquidity is high. For example, Across maintains the largest median total solver liquidity, but also exhibits the longest median refund time, 
This period is the \emph{attack window} for a liquidity exhaustion attack.

\featheader{Intent Value Distribution} While Mayan and Across have comparable median intent values, deBridge processes significantly larger transfers, amplifying the impact of individual large intents on solver liquidity.

%Also show Ethereum as the main profitability hub. Number of intents for each dst blockchain to justify only studying Ethereum.

\begin{figure}[ht]
    \centering
    \includegraphics[width=0.5\textwidth]{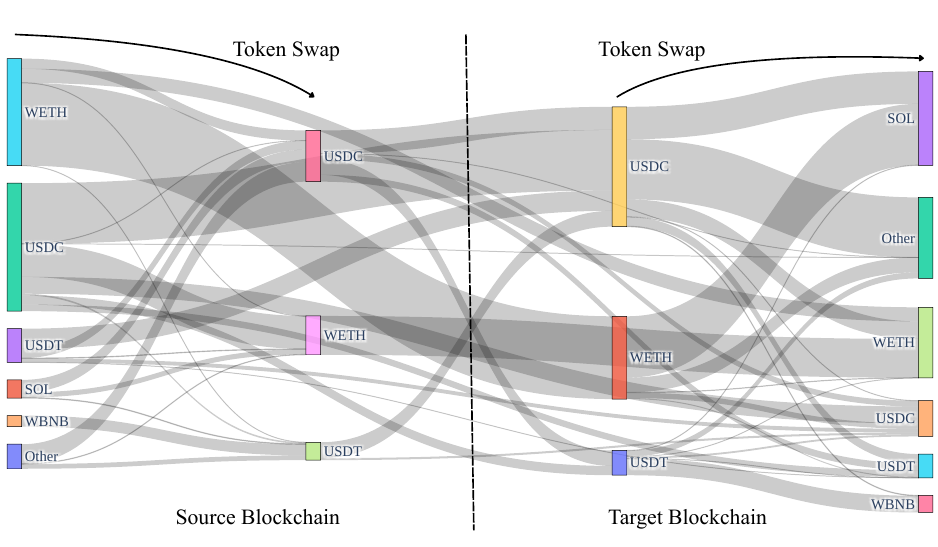}
    \caption{Simplified token routing: low-liquidity tokens are converted to a small set of high-liquidity tokens (e.g., ETH, USDC, USDT) for efficient solver balance management. If the user requests a different token on the destination chain, these are optionally swapped accordingly.}

    \label{fig: token_swaps}
\end{figure}

% Only needed liquidity of a couple of tokens. SHOW PLOT.
\featheader{Token Balances} Solvers only need to maintain balances in a small set of highly liquid tokens. Although users may specify a wide variety of assets to receive on the destination chain, solvers typically fulfill intents using a limited set of base assets (e.g., ETH, USDC, and USDT) and perform token swaps to obtain the requested tokens (Figure~\ref{fig: token_swaps} depicts this flow for Mayan Swift, while Across and deBridge follow similar patterns). This information allows to efficiently calculate the total liquidity available for each solver at each timestamp.

%Figure~\ref{fig: token_swaps} illustrates that the vast majority of destination-side transfers and swaps ultimately originate from these core tokens. %This behavior significantly simplifies solver balance management, as effective liquidity can be approximated by tracking balances of only a few assets rather than the full universe of tokens supported by the protocol.

\section{Liquidity Exhaustion Attacks}
\label{sec: liq_exhaustion_attacks}

\begin{figure*}[t]
    \centering
    \includegraphics[width=0.95\linewidth]{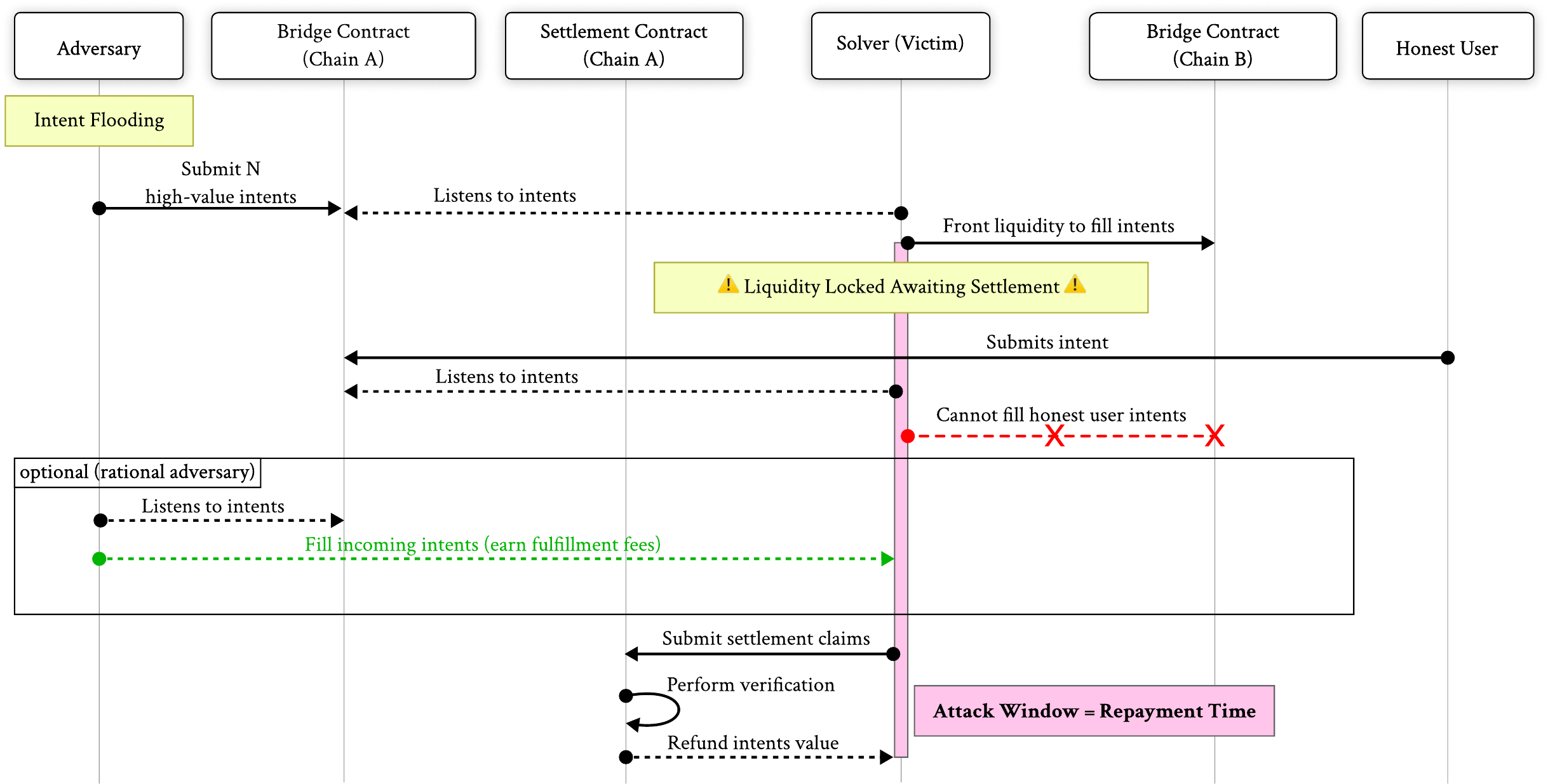}
    \caption{Sequence diagram of a liquidity exhaustion attack, showing the interaction between attacker, solvers, and protocol.}
    \label{fig: attack_sequence}
\end{figure*}

We introduce \emph{liquidity exhaustion attacks}, a previously unexplored class of attacks against intent-based cross-chain protocols. These attacks arise from a fundamental design property of such systems: solvers must front liquidity to fulfill intents immediately, while reimbursement occurs only after an asynchronous and variable settlement delay. An attacker $\mathbb{A}$ exploits a protocol by submitting carefully timed large-value intents that temporarily exhaust solvers' liquidity (\emph{Intent Flooding} stage), causing solvers to become unable to bid in subsequent intents and consequently to fulfill subsequent legitimate requests. %Unlike traditional denial-of-service attacks that target computational or network resources, liquidity exhaustion attacks target the economic capacity of solvers.
Figure~\ref{fig: attack_sequence} contains a sequence diagram showing the sequence of steps to induce the attack.

% move to motivation??
%An attacker may want to study the possibility of profiting from these strategies. A solver may want to study the risks of having specific amounts of liquidity available and choose intent-fulfillment strategies accordingly. For solvers, there is an additional focus point: the refund process. Whenever possible, if solvers ask for refunds less times they pay less in transaction fees but remain more time with liquidity shortage. On the other hand, if solvers as for refunds more times they will spend more on fees but being near to maximum liquidity in the majority of time. The additional objective of the framework is to calculate the breaking point, where solvers can minimize the costs of getting refunds without having the susceptibility of being a target of a DoS attack.

\textbf{Attack Goal:} Temporarily drain the liquidity of solvers in a cross-chain protocol to 1) maximize the number of intents fulfilled by an attacker in order to maximize profit; or 2) break the availability of the protocol, recreating a denial of service attack through liquidity exhaustion.

\subsection{Adversarial Model}\label{subsec: adversarial_model}
We model an adversary $\mathbb{A}$ who can generate arbitrary cross-chain intents and strategically choose their timing based on publicly observable information such as solver liquidity, repayment delays, and historical fill behavior. $\mathbb{A}$ may pursue two goals: (i) a \textbf{byzantine} (malicious) strategy that seeks to degrade protocol liveness or harm specific solvers even at financial loss; or (ii) a \textbf{rational} and profit-driven strategy that exploits temporary liquidity imbalances to extract financial gain. $\mathbb{A}$ can submit intents on any supported blockchain, target one solver or multiple simultaneously, and possesses enough capital to execute the attack but cannot interfere with settlement logic, or compromise other solvers or protocol configurations. Finally, a rational adversary needs to be an eligible solver in each protocol.

\begin{comment}
\begin{table}[h]
\centering
\small
\begin{tabular}{lccc}
\toprule
\textbf{Capability} & \textbf{Baseline} & \textbf{Targeted} & \textbf{Byzantine} \\
\midrule
Observe solver liquidity & \checkmark & \checkmark & \checkmark \\
Observe solver participation & -- & \checkmark & -- \\
Inject adversarial intents & \checkmark & \checkmark & \checkmark \\
Economic motivation & \checkmark & \checkmark & -- \\
\bottomrule
\end{tabular}
\caption{Summary of attacker capabilities across models.}
\label{tab:threat_model}
\end{table}
\end{comment}

\subsection{Attack Models}
\label{subsec: attack_models}

We index intents by $i$. For each historical intent, we observe the following quantities from on-chain data:
\begin{itemize}\setlength\itemsep{-0.2em}
    \item $V_i$: value transacted in intent $i$,
    \item $p_i$: solver profit margin for fulfilling intent $i$,
    \item $F^{\text{prot}}_i$: effective protocol fee rate (usually composed of a percentage of the $V_i$ plus a fixed fee) charged for intent $i$,
    \item $g_i$: transaction fee incurred when fulfilling intent $i$,
    \item $C_i^{\text{auction}}$: transaction fees incurred when bidding in the auction for intent $i$,
    \item $t_i$: timestamp at which intent $i$ is created.
\end{itemize}

Let $L(t)$ denote the total liquidity available by all solvers at timestamp $t$.

\subsubsection{Liquidity Exhaustion and Attack Induction Cost}

We model liquidity exhaustion abstractly via an exhaustion parameter $\alpha \in [0,1]$, which represents the fraction of competing solver liquidity that becomes unavailable during an attack window. This parameter is not optimized analytically; instead, it serves as a conceptual abstraction that captures varying degrees of liquidity exhaustion observed in our simulations (mainly in the targeted attack strategy in Section~\ref{sec: targeted_attack}).

The cost incurred by $\mathbb{A}$ to exhaust liquidity at time $t$ is modeled as:

\begin{equation}
\label{eq: cost_induction}
C_\text{induction}(\alpha, t) = \alpha \cdot L(t) \cdot F^{\text{prot}} + C^{\text{gas}}_\text{induction}(t)
\end{equation}
where $\alpha \cdot L(t) \cdot F^{\text{prot}}$ represents the capital required to temporarily remove competing liquidity, $C^{\text{gas}}_\text{induction}(t)$ captures the blockchain transaction fees paid during in the \emph{Intent Flooding} stage. This expression reflects that inducing greater liquidity exhaustion requires proportionally more capital and transaction fees.

\subsubsection{Cost to Fill Intents}

The attacker also incurs costs to actually fulfill intents:

\begin{equation}
\label{eq: cost_fill_intents}
C_\text{fill\_intents} = \sum_i \Big( g_i \;+\; C_i^{\text{auction}} \Big)
\end{equation}
where $g_i$ is the gas fee and $C_i^{\text{auction}}$ is the auction bidding cost for intent $i$.

\subsubsection{Attack Revenue}

When the attacker successfully fulfills intent $i$, they earn:

\begin{equation}
R_i = V_i \cdot p_i% \cdot \mathbb{I}_i
\end{equation}
%
%where $\mathbb{I}_i$ is an indicator variable equal to 1 if the attacker fulfills intent $i$ and 0 otherwise. Empirically, $\mathbb{I}_i$ is determined via replay-based simulations rather than a probabilistic model.

Total attack revenue is then given by:

\begin{equation}
r_\text{attack} = \sum_i R_i
\end{equation}

\subsubsection{Net Profit}

Finally, the attacker’s net profit over the attack window is:

\begin{equation}
\label{eq: overall_profit}
\text{Profit} = r_\text{attack} \;-\; C_\text{induction} \;-\; C_\text{fill\_intents} \;-\; \epsilon
\end{equation}
where $\epsilon$ accounts for an additional potential cost of adverse price effects due to large token movements, slippage, swap fees, opportunity costs, etc.

\section{Data Collection and Analysis}
\label{sec: data_collection_and_analysis}

In this section, we explain our data extraction process, the interval under analysis, and present a first analysis that summarizes the key vulnerabilities that, theoretically, make the liquidity exhaustion attacks described in Section~\ref{sec: liq_exhaustion_attacks} possible.

\subsection{Dataset Overview}
\label{sec:dataset}

Our study analyzes three intent-based cross-chain protocols: Mayan Swift, Across, and deBridge. These protocols rank among the largest intent-based bridges by volume and together represent a substantial fraction of cross-chain activity~\cite{defillama}. Several other systems rely on one or more of these protocols, making them representative of the broader intent-based bridging ecosystem. The dataset covers the period from June 1, 2025, to November 1, 2025, and includes blockchains accounting for over 90\% of the total value transacted by each protocol (including Solana, Arbitrum, Ethereum, Base, Polygon, BNB, Unichain, Optimism, and Avalanche).

We construct a unified dataset (the schema used is in Appendix \ref{appendix: data_model}) by extracting on-chain event data from each protocol's smart contracts. Solver liquidity is reconstructed from on-chain token transfer events from high-liquidity tokens (such as ETH, USDC, USDT, WBNB, WBTC -- cf. Figure~\ref{fig: token_swaps}) and balance queries to public RPC endpoints, while transactions on different blockchains are merged using protocol-specific identifiers. Token prices are obtained using Alchemy's Token Prices API \cite{alchemy} with daily granularity, and solver profitability, protocol fees, and fulfillment latency are derived off-chain from the reconstructed traces. The dataset and code will be made publicly available.

\subsection{Vulnerabilities Identified}
\label{sec: vulnerabilities_identified}

Our analysis reveals structural patterns making intent-based protocols vulnerable to liquidity exhaustion attacks.

\begin{figure}[h]
    \centering
    \includegraphics[width=0.47\textwidth]{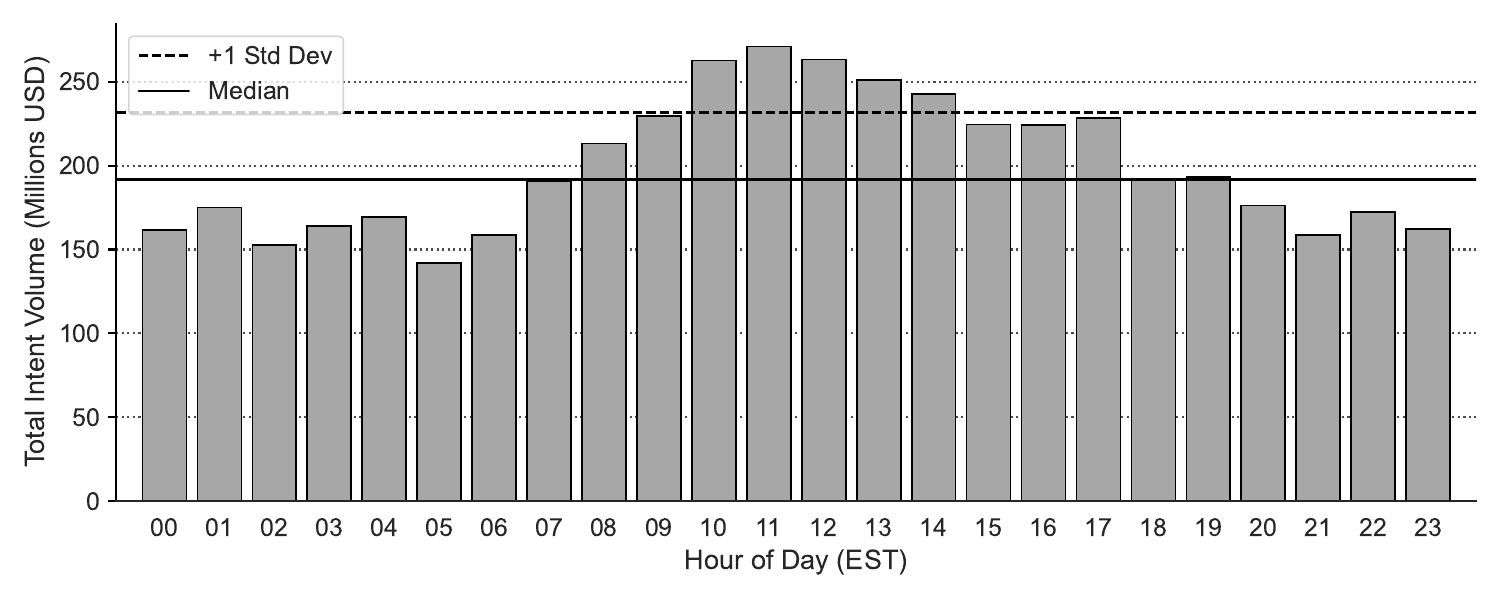}
    \caption{Hourly distribution of intent volume for deBridge, showing strong temporal concentration of value transacted between 9am-2pm EST.}
    \label{fig: hourly_intent_volume_debridge}
\end{figure}

\featheader{Temporal Concentration of Demand}
Intent activity is highly time-dependent. In all protocols analyzed, the majority of the total value transacted occurs within a four-hour window (10am -- 2pm EST). Figure~\ref{fig: hourly_intent_volume_debridge} illustrates this concentration for deBridge, with similar patterns observed for Mayan and Across. These periods concentrate liquidity demand, increasing the likelihood that large intents temporarily exhaust available liquidity.

\featheader{Asymmetric Profitability Across Chains}
Solver profitability varies significantly across blockchain pairs. Ethereum consistently yields the highest per-intent solver profits across all protocols, while transferring from Solana exhibits the lowest user-paid fees for transfers into Ethereum. More specifically, token transfers from Solana are 30 -- 55\% cheaper for users while token transfers to Ethereum yield 2 -- 4x higher solver profits compared to other paths.%Therefore, in our simulations, we use Solana to Ethereum intents as base.

% REWRITE WITHOUT HEATMAP INFO
%Figure~\ref{fig: profitability_heatmap_debridge}, in Appendix highlights regions where this asymmetry makes liquidity exhaustion attacks economically favorable. For this reason, we focus our simulations on Solana-to-Ethereum transfers for Mayan and deBridge, and Base-to-Ethereum transfers for Across.

%\featheader{Solver Market Concentration}
%Solver participation is mainly unevenly distributed.

% REWRITE WITHOUT HHI INFO
%Figure~\ref{fig:hhi} shows that some protocols exhibit extreme solver concentration, with a single solver responsible for over \textbf{90\% of fulfills}. Such concentration creates effective single points of failure, where draining the liquidity of a small number of solvers can disproportionately affect protocol-wide availability.

\begin{comment}
    
\begin{figure}
    \centering
    \includegraphics[width=0.47\textwidth]{figures/profitability_heatmap_debridge.png}
    \caption{Solver profitability heatmap for deBridge, highlighting blockchain pairs where liquidity exhaustion attacks are most economically favorable.}
    \label{fig:profitability_heatmap_debridge}
\end{figure}

\begin{figure}
    \centering
    \includegraphics[width=0.47\textwidth]{figures/hhi.png}
    \caption{Solver market concentration measured using the Herfindahl--Hirschman Index (HHI), indicating solver centralization across protocols.}
    \label{fig:hhi}
\end{figure}
\end{comment}

\begin{figure}[h]
    \centering
    \includegraphics[width=\linewidth]{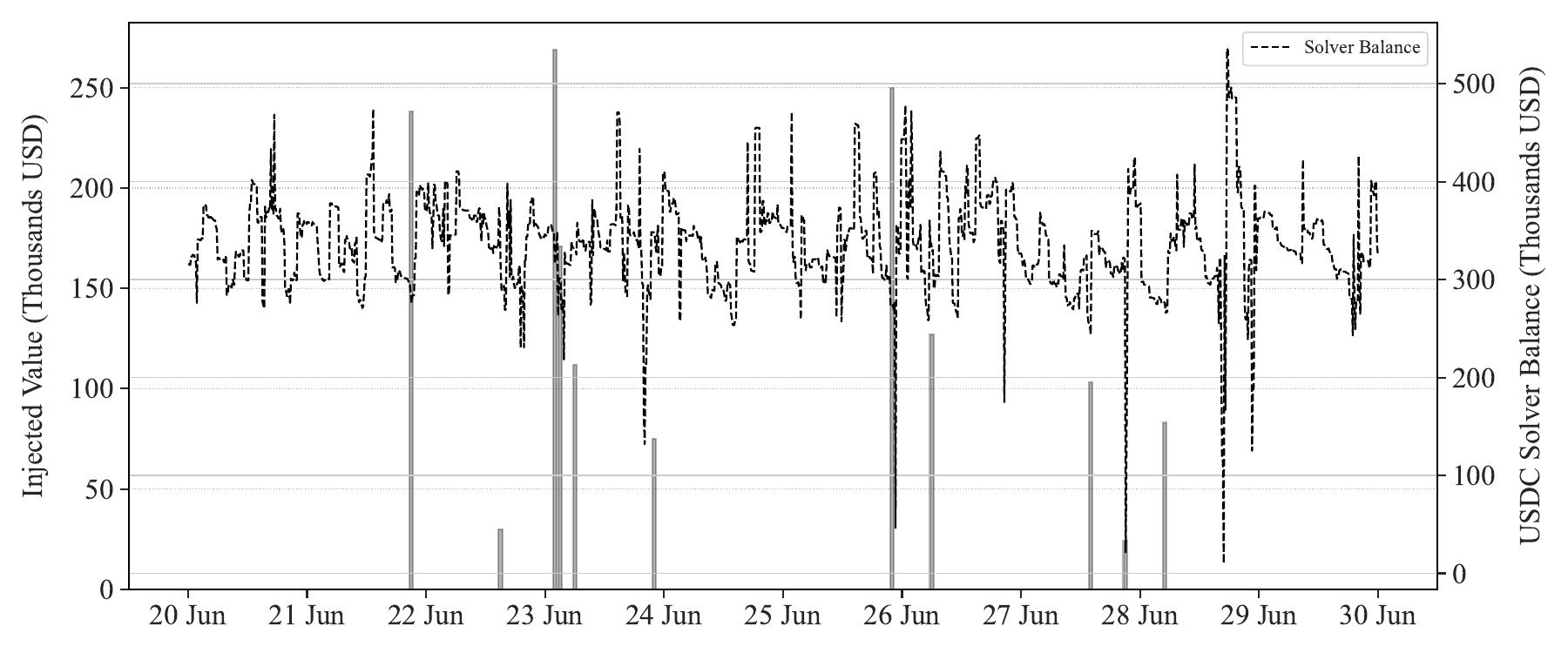}
    \caption{Hourly liquidity injections (bars) versus solver balance (dashed line) for Mayan Swift’s top solver between June 20 and June 30, 2025. The data reveals no automatic mechanism of injection of liquidity once liquidity goes below a certain threshold.}
    \label{fig: liquidity_injections_native_mayan}
\end{figure}

\featheader{Solver Rebalancing and Liquidity Injections}
We examine whether solver liquidity is actively rebalanced in response to low balances or follows a systematic pattern. %We define a \emph{liquidity injection} as an inbound token transfer to a solver address and analyze inbound transfers for the most active solvers in all protocols. Solver balances are reconstructed from on-chain data and aligned with observed injection events.
Figure~\ref{fig: liquidity_injections_native_mayan} shows an example period for the most active Mayan Swift solver. In the entire dataset, we find no evidence of automated or reactive rebalancing: liquidity injections do not occur at regular intervals, are not triggered by balances crossing consistent thresholds, and do not exhibit stable or proportional sizes. Instead, injections appear sporadic and weakly correlated with short-term liquidity shortage, suggesting that transient liquidity dips can persist long enough to be exploited by opportunistic attackers. 

\begin{figure}[ht]
    \centering
    \includegraphics[width=0.47\textwidth]{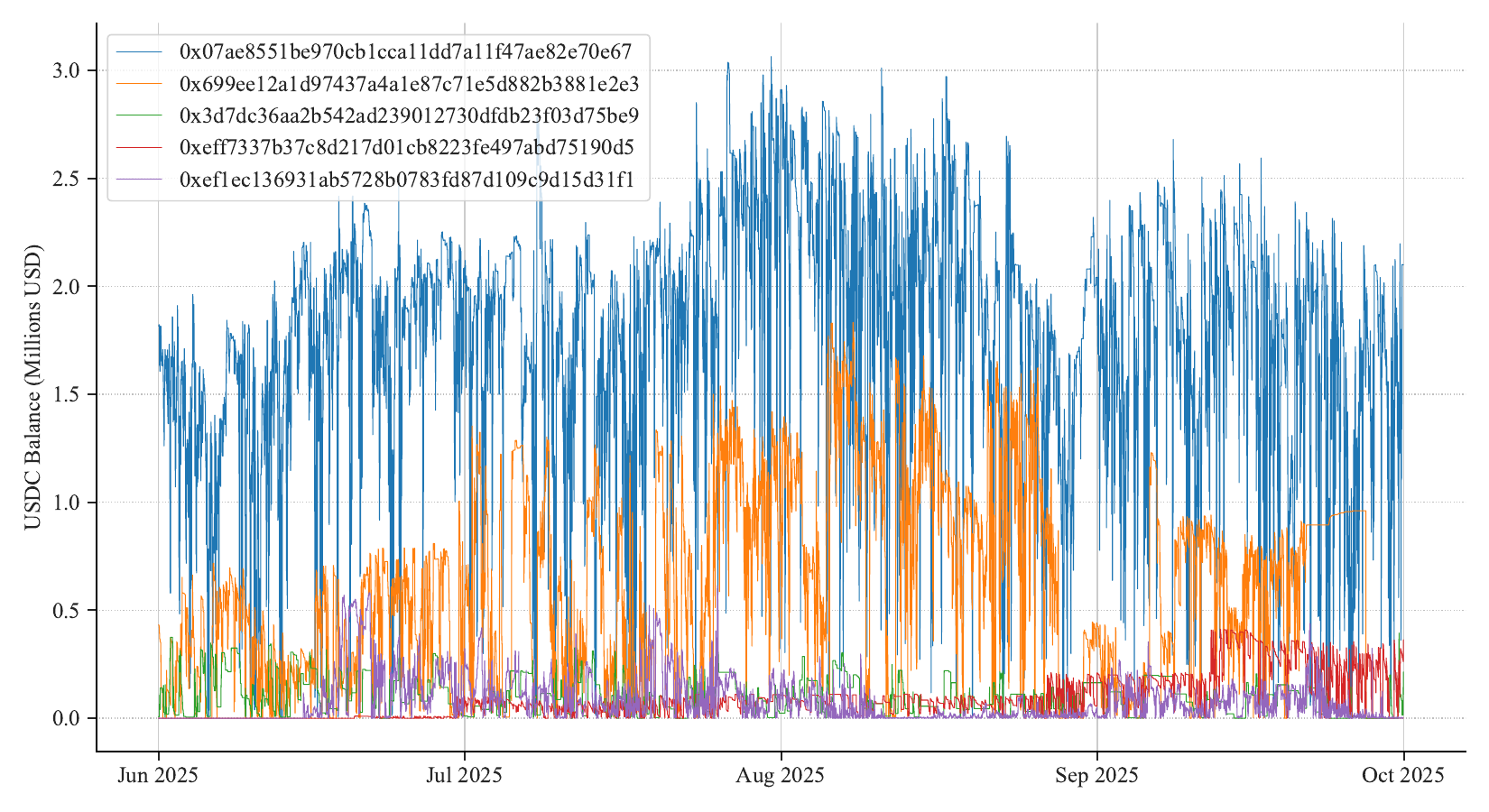}
    \caption{USDC balances of Across solvers on Ethereum, showing volatile liquidity and recurring depletion events.}
    \label{fig: solvers_over_time_across_USDC_ethereum}
\end{figure}

\featheader{Liquidity Volatility and Attack Windows}
Figure~\ref{fig: solvers_over_time_across_USDC_ethereum} depicts the volatility of solver balances over time. The volatility is created by the fulfillment and repayment of intents. The recurrent dips create recurring attack windows that can be exploited when combined with delayed settlement and limited liquidity injections.

\begin{comment}
\begin{figure}[h]
    \centering
    \includegraphics[width=0.47\textwidth]{figures/solvers_over_time_mayan_native_ethereum.png}
    \caption{Native ETH balances of Mayan Swift solvers on Ethereum, illustrating short-lived liquidity dips that create attack windows.}
    \label{fig:solvers_over_time_mayan_native_ethereum}
\end{figure}
\end{comment}

\begin{findingbox}
\textbf{Main Insights.} Intent-based cross-chain protocols concentrate demand in predictable time windows, exhibit asymmetric solver profitability across chains, rely on highly centralized solver markets, and experience volatile solver liquidity. These characteristics enable economically feasible liquidity exhaustion attacks through adversarial timing.
\end{findingbox}

\section{Attack Simulation Framework}
\label{sec:simulation_framework}

To evaluate liquidity exhaustion attacks, we create a replay-based simulation framework that injects adversarial intents into historical cross-chain transaction traces. The goal of the framework is to determine whether, and under which conditions, an attacker can temporarily exhaust solver liquidity and disrupt intent fulfillment in real protocol deployments. The framework is fully parameterized and supports different blockchains, bridge protocols, and economic assumptions. The implementation will be made publicly available.

\subsection{Design Rationale}
We adopt replay-based simulations to avoid assumptions about solver behavior, bidding strategies, or liquidity management that are not publicly observable. Grounding simulations in historical execution traces, ensures that attack outcomes reflect real protocol usage patterns, intent value distributions, and solver participation, rather than synthetic behavior. Most importantly, the costs incurred by solvers when fulfilling intents ($C_{fill\_intents}(t)$ and $\epsilon(t)$ in Equations~\ref{eq: cost_fill_intents} and~\ref{eq: overall_profit}, respectively) are exactly the same values that we use to simulate an attacker fulfilling exactly the same intent at that time.

Each simulation is instantiated by a configuration specifying the source blockchain, destination blockchain, bridge protocol, attack window duration, and maximum adversarial intent value. Unless explicitly stated otherwise, simulations use real protocol parameters (i.e., the real values of solver profitability, protocol fees, and user costs). For example, the configuration \texttt{SRC\_BLOCKCHAIN=solana}, \texttt{DST\_BLOCKCHAIN=ethereum}, and \texttt{BRIDGE=mayan} replays historical Mayan Swift transfers from Solana to Ethereum without modifying solver margins or user fees. In our experiments, we range \texttt{ATTACK\_WINDOW} from 300 to 1000 seconds (we discuss these choices in~\ref{sec:discussion}). Additionally, using \texttt{MAX\_TX\_VALUE=10000} restricts adversarial intents to a maximum value of \$10{,}000 each (this value will tell how many intents $\mathbb{A}$ needs to create in the flooding phase). Finally, \texttt{VOLUME\_MULTIPLIER=1} will force the attacker to use a window multiplier of 1 to preserve the original transaction volume.

%For a given simulation, the attacker injects adversarial intents at a chosen timestamp $t_s$. The protocol's subsequent behavior over the attack window is replayed using historical data to determine which intents would be fulfilled or fail due to insufficient solver liquidity. Adversarial intents are treated identically to historical intents, except that their submission timing and value bounds are controlled by the attacker.

\subsection{Simulated vs.\ Fixed Parameters}
Our evaluation distinguishes between parameters fixed to historical on-chain values and parameters varied as controlled stress tests. In all tables, the parameters set to their observed values are labeled as ``Real'', while the simulated parameters report the effective value used. The simulated values are not arbitrary: whenever a parameter is varied, we reuse the values observed in one of the other protocols (in Table~\ref{tab: protocols-summary}), allowing cross-protocol stress testing under realistic operating conditions. Transaction volume is varied via a volume multiplier, where a value of \texttt{1} corresponds to the real historical volume, and larger values scale total volume to match that of another protocol in our dataset, allowing us to assess protocol behavior under demand levels realized elsewhere in the ecosystem.

\subsection{Attack Cost and Outcome Estimation}

Since solver implementations and liquidity management strategies are not publicly observable, we treat these components using a black-box model. We compute the volume and value of historical intents that fail to be fulfilled during the attack window due to liquidity exhaustion. This yields conservative estimates of attack profitability: \textbf{in a live deployment, reduced solver competition would likely decrease bidding pressure in auctions and yield lower attack costs relative to our estimates}. In contrast, attack costs are estimated using protocol fee trends and historical gas costs observed at $t_s$, without assuming any favorable pricing for $\mathbb{A}$.

Solver liquidity fluctuates substantially over time due to ongoing fulfillments, delayed refunds, and rebalancing behavior (cf. Figure~\ref{fig: solvers_over_time_across_USDC_ethereum}). As a result, identical attack parameters can produce markedly different outcomes depending on the timing. To account for this variability, all reported results aggregate identical attack configurations across multiple attack timestamps within the analysis period, capturing both typical and tail behavior (using the 90th percentile), to capture the heavy right-skewed distributions.

%\subsubsection{Framework Inputs}

%Src and Dst Blockchain for calculation of the protocol fees of the intents created.

%Solver's balances of the tokens supported (e.g., USDC, USDT, ETH, DAI, WBTC).

\section{Baseline Strategy: Median-Deviation Liquidity Trigger}
\label{sec: baseline_strategy}

In our baseline strategy, we exploit the naturally occurring liquidity fluctuations. Instead of attempting to forecast future profitability or solver behavior, $\mathbb{A}$ opportunistically triggers attacks during periods when solver liquidity is unusually low relative to historical levels.

\subsection{Attack Model}

For each solver $s$, $\mathbb{A}$ observes its historical liquidity time series $L_s(t)$. At each timestamp $t$, $\mathbb{A}$ computes the historical median liquidity $\tilde{L}_s$ and standard deviation $\sigma_s$. We use the median rather than the mean to mitigate the impact of huge liquidity injections and intent values that introduce heavy-tailed variability. The objective is to identify periods of low liquidity that create adversarial opportunities. Inspired by median-deviation outlier detection methods, $\mathbb{A}$ triggers an attack whenever:

\begin{equation}
L_s(t) < \tilde{L}_s - k \cdot \sigma_s,
\end{equation}
where $k \in \mathbb{N}$ controls the selectivity of the trigger. Lower values of $k$ result in more frequent attacks, while higher values isolate rarer but more severe liquidity shortages: with $k = 0$, attacks trigger whenever liquidity falls below the historical median; with $k = 1$, attacks trigger whenever liquidity falls below the historical median minus a standard deviation; and so on. This strategy requires no forecasting methods and exploits periods with severe liquidity shortage during which the capital required to exhaust the liquidity is reduced.

%\subsection{Experimental Setup}
%For each protocol, we vary the deviation threshold $k \in \{0,1,2,3\}$, the attack window duration (300s, 600s, 1000s), and protocol-specific parameters including solver profitability, protocol fees, maximum adversarial intent value, and the protocol intent volume.

%In the main body, we present a representative configuration with $k=0$ and a 1000~s attack window (Table~\ref{tab: results}); full sensitivity results are reported in Appendix~XXX.

\subsection{Optimal Deviation Threshold (k-values)}\label{subsec: k_values}

%\mpc{falta definir o que é o deviation threshold e (talvez) o que é que tem que ver com Attack Frequency (k-values), que é o tema da secção. É o mesmo k que aparece bom fim da secção anterior? Se sim, porque é que lá não tinha esta designação? Se não, é melhor não usar a mesma letra pois gera confusão.}
For each protocol, we vary the deviation threshold $k \in \{0,1,2,3\}$. For $k=3$, only a small number of extreme liquidity shortages remain, and only for Mayan Swift. For $k > 3$, there are no attack instances returned, i.e., the liquidity is never that low.

Tables~\ref{tab: k_sensitivity_1000} and~\ref{tab: k_sensitivity_300} summarize the attack outcomes across opposite attack durations (1000s and 300s, respectively), without any simulated parameters. The column \texttt{Pr[Profit]} shows the percentage of attacks that yielded (positive) profit, to avoid a few isolated intents to boost the overall net profit.
%\mpc{a primeira frase está confusa pois pareces estar a falar do k e depois aparece entre parênteses 1000s and 300s que não podem ser valores do k; são valores de quê?}
%\andre{tem razão, estavam trocados!}

%For $k=0$, attacks trigger whenever liquidity falls below the historical median, producing thousands of attack windows (e.g., 1,833 for Mayan and 2,782 for deBridge with a 1000~s window). Increasing $k$ sharply reduces the number of eligible windows; 

The results show that, for an attack window of 1000s, deBridge is clearly vulnerable for every $k \in \{1,2\}$. For $k=0$, the attack is performed plenty of times when the liquidity is not that low, which increases the total cost of performing the attack and returns financial loss. Interestingly, with $k=3$, the attack is performed fewer times (fulfilling only $\approx92k$ compared to $\approx107k$ transactions in $k=1$), but the Pr[profit] is 100\%. This means that all attack instances returned profit, but the total net profit is lower than $k=1$, indicating a reliable attack in the long term. In Mayan Swift, there are a few instances where attacks are feasible at $k=0$ and $k=1$, becoming effectively infeasible for higher thresholds. However, in the long term, the attack is not reliable because the majority of instances return a financial loss. From our investigation, there are specific timestamps yielding profit due to high-value intents being transacted at that point, which indicates that the security of the protocol against a liquidity exhaustion attack is somehow dependent on protocol-external variables, such as the user behavior. In order to understand the exact impact of different variables in making the attack possible in some instances, we study the impact of different parameters in the simulations of Section~\ref{subsec: parameter_simulation}. Across shows no meaningful attack opportunities for any tested $k$, even under the most permissive trigger, indicating that it is sufficiently stable to prevent opportunistic attacks under real operating conditions.

\input{tables/k-sensitivity}

\begin{figure}[h]
    \centering
    \includegraphics[width=\linewidth]{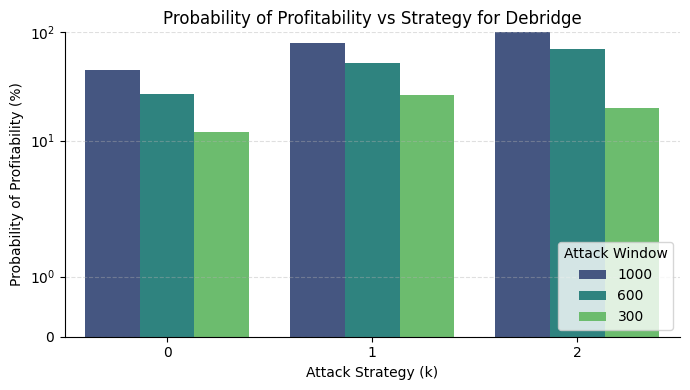}
    \caption{Probability that an attack window yields positive net profit for deBridge under the median-deviation strategy, as a function of deviation threshold $k$.}
    \label{fig: prob_profitability_over_k_debridge}
\end{figure}

\begin{figure}[h]
    \centering
    \includegraphics[width=\linewidth]{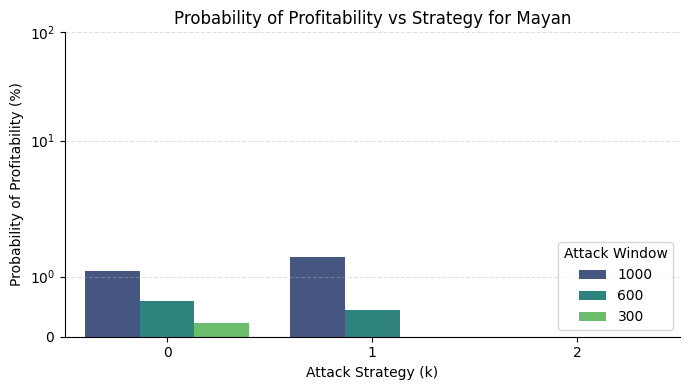}
    \caption{Probability that an attack window yields positive net profit for Mayan Swift under the median-deviation strategy, as a function of deviation threshold $k$.}
    \label{fig: prob_profitability_over_k_mayan}
\end{figure}

In the same timestamp, reducing the attack window from 1000~s to 600~s or 300~s decreases the volume that can be captured by the attacker, which decreases the revenue and consequently the profit. Figures~\ref{fig: prob_profitability_over_k_debridge} and~\ref{fig: prob_profitability_over_k_mayan} depict the expected profitability of attacks as a variable of the attack window and the value of $k$. In deBridge, for $k = 2$ (the most reliable configuration per Table~\ref{tab: k_sensitivity_1000}), reducing from 1000~s to 600~s qualitatively does not change the results. However, reducing the window to 300~s (cf. Table~\ref{tab: k_sensitivity_300}) largely eliminates profitable opportunities, indicating that short-lived liquidity exhaustion alone is insufficient to sustain economically viable attacks.

\begin{findingbox}
\textbf{Main Insights.} Under real conditions, a rational attacker using the median-deviation strategy is able to reliably extract profit in deBridge in the long term. An attack targeting Across is completely unfeasible with no instance returning profit due to the very low solver profitability and massive liquidity. In Mayan Swift, the attacker could profit in specific timestamps due to high-value intents, indicating that the security of the protocol is dependent on protocol-external variables.
\end{findingbox}

\input{tables/main-body-results-dos}

\subsection{Parameter Simulation}\label{subsec: parameter_simulation}

For every $k$ and attack window duration, we simulate liquidity exhaustion attacks varying the solver profitability, the protocol fee charged, the maximum value of each created intent, and the volume multiplier. Table~\ref{tab: results} summarizes the results for $k=1$. The remaining results will be available in our open-source implementation.

\featheader{Mayan Swift}
For Mayan, the baseline strategy is generally unprofitable, but becomes profitable in a few instances under specific parameter combinations, mainly when solver profitability or transaction volume increases. This highlights a key trade-off when designing intent-based systems: higher solver margins may incentivize participation of more solvers, but also increase susceptibility to such an attack. When increasing the solver profitability from 0.381\% to 1.129\% (deBridge's value), the attack profitability becomes a concern with $\approx 49\%$ of instances returning profit, where the 90th percentile is a net profit of \$852. Also, the fact that over 20\% of the attacks performed, considering a 5.61x increase in the intent volume (simulating Across's volume), demonstrates that the risk also depends on external usage patterns rather than protocol logic alone. This conclusion is in line with the findings obtained in Section~\ref{subsec: k_values}.

\featheader{deBridge}
deBridge is the most susceptible protocol under the baseline strategy. Across a wide range of parameters, simulations yield positive expected net profit, albeit with high variance driven by frequent high-value intents. Profitability persists even for moderate liquidity deviations, indicating that common liquidity dips combined with relatively high solver margins enable opportunistic attacks. The only instance where the protocol is not subject to such attacks is when the solver profitability is simulated to be equal to Across's, as low as 0.018\% (compared with deBridge's 1.129\%).

\featheader{Across}
Across consistently yields a negative net profit across all evaluated configurations. Extremely low solver margins dominate attacker economics, yielding rational attacks unprofitable even when massive transaction volume is captured. Across also maintains substantially higher total liquidity (over \$8 million, which is drastically different from Mayan and deBridge -- cf. Figure~\ref{tab: protocols-summary}), which sharply increases the cost of the attacks. Keeping solver profitability very low is an effective defense against economically motivated attacks, but it risks discouraging participation by making solver operation less attractive relative to other opportunities in the ecosystem.%\andre{if there's time I'll add some values of solver profitability to give a bit more strenght to the argument.}

\begin{comment}
\featheader{Sensitivity to Liquidity}
Figure~\ref{fig: profitability_vs_liquidity_exhaustion} shows that profitability increases non-linearly beyond protocol-specific liquidity thresholds. deBridge exhibits the steepest transition, followed by Mayan, while Across requires substantially higher exhaustion levels before attacks become viable.

\begin{figure}[h]
    \centering
    \includegraphics[width=\linewidth]{figures/proft_vs_liquidity_variation.png}
    \caption{Probability that an attack yields positive net profit for each bridge under the median-deviation strategy, as a function of the total liquidity available $L_s(t)$.}
    \label{fig: profitability_vs_liquidity_exhaustion}
\end{figure}
\end{comment}

\input{tables/results-byzantine}

\begin{findingbox}
\textbf{Main Insights.} With volatile liquidity and sufficiently high solver profitability, naturally occurring liquidity dips alone can enable economically viable attacks using the baseline strategy. Conversely, consistently low solver margins act as the strongest defense strategy. These findings motivate more selective attacker strategies, which we explore next.
\end{findingbox}

\section{Byzantine Adversary: Availability Impact}
\label{sec: byzantine_adversary}

% repeated from Adversary Models
%We consider a worst-case byzantine adversary $\mathbb{A}$ capable of temporarily suppressing the participation of all solvers in an intent-based cross-chain protocol during a bounded attack window.
Unlike rational adversaries that seek to optimize profit subject to liquidity constraints, a byzantine attacker $\mathbb{A}$ is not limited by economic incentives and aims to maximize availability disruption. $\mathbb{A}$ is assumed to prevent solvers from fulfilling intents for the duration of the attack window, resulting in failed or delayed intent fulfillments. For each protocol and attack window duration, we simulate thousands of attack instances at different historical timestamps. Each instance aggregates the intents that would fail to be fulfilled within the window, producing per-window metrics such as the number of failed intents, the median failed intent value, and the economic impact in terms of missed solver profit and protocol fees.

Table~\ref{tab: byzantine_p90} reports distributional summaries across all simulated attack windows, including medians and 90th percentile, to capture both typical and worst-case behavior.

\begin{comment}
\begin{figure}[ht]
    \centering
    \includegraphics[width=\linewidth]{figures/byzantine_across.png}
    \caption{Caption}
    \label{fig:byzantine_across}
\end{figure}
\end{comment}

\subsection{Availability and Economic Impact}
Failed intent values exhibit heavy-tailed behavior: median values remain modest between the \$74 and \$200, while 90th percentiles range from $\approx\$223$ to $\approx\$1200$. In order to show the high variability of the value of intents processed by protocols, we show the standard deviation of every attack window, and show that the 90-percentile ranges from \$5k to \$43k. This means that, even though the majority of the failed intents are pretty low in value and the number of failed intents is not substantial, there are users trying to bridge massive amounts of funds, and these are impacted by such an attack.

Although $\mathbb{A}$ does not profit, the attack imposes opportunity costs on solvers and protocol operators through missed fees and profit. Solvers may lose up to almost \$1000 in missed solver profitability in deBridge, \$160 in Across in 1{,}000 seconds. Mayan Swift has the most modest number of transactions and value transacted, which means that solver profitability rarely goes beyond \$100 even in a 1000s window. Similarly, protocols can lose up to $\approx \$187$, $\approx \$285$, and $\approx \$26$ in protocol fees per attack window in deBridge, Across, and Mayan, respectively. In terms of attack execution costs, the median is approximately \$2.8k for Across, \$239 for Debridge, and \$339 for Mayan, largely independent of duration, since the full liquidity must be drained in every case.

\subsection{Direct vs. Indirect Impact}

\begin{figure}[ht]
    \centering
    \includegraphics[width=\linewidth]{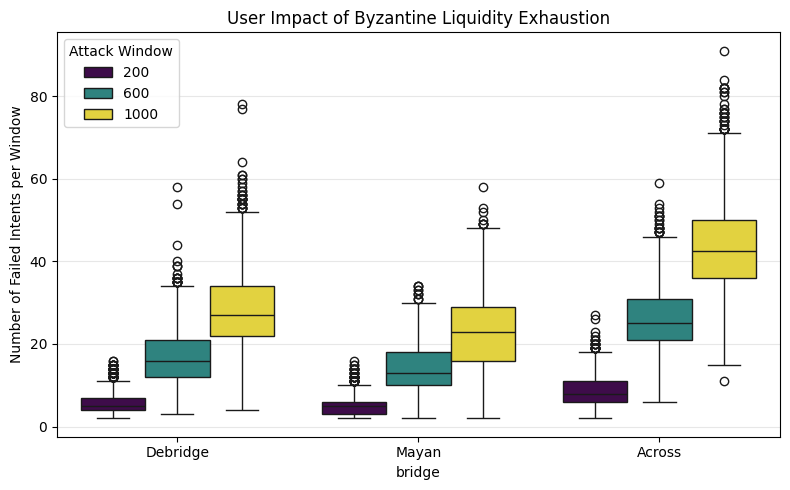}
    \caption{Impact of the liquidity exhaustion attack in terms of the number of user intents rejected by Across protocol.}
    \label{fig: user_impact_byzantine}
\end{figure}

Liquidity exhaustion attacks have both direct and indirect consequences for users and protocols. The direct impact on users is immediate: the number of failed intents increases monotonically with the duration of the attack (cf.\ Figure~\ref{fig: user_impact_byzantine}). Across the evaluated protocols, the number of failed intents ranges from only a few instances in Mayan Swift to nearly one hundred in Across, which processes the highest transaction volume. For affected users, these failures translate into delayed or reverted cross-chain transfers, directly degrading the user experience.

At the protocol level, as shown by Table~\ref{tab: byzantine_p90}, the direct financial impact is limited. During attack windows, protocols miss a small amount of fee revenue due to unfulfilled intents. However, these losses are negligible compared to the total value processed by each protocol, which is on the order of millions or billions of USD~\cite{augusto2025xchaindatagen}. As a result, liquidity exhaustion attacks do not pose an existential financial threat solely through lost fees. The most significant risk lies in the indirect impact. Cross-chain bridges increasingly serve as infrastructure for aggregation services, including DEX aggregators and bridge aggregators~\cite{10664224}, which rely on high availability and correct fulfillment of transactions to route user transactions (according to our data, only around 25.23\% of the intents processed by Across do not come from these third-party services). Even short-lived availability failures can cause these aggregation services to route traffic away from a protocol due to perceived unreliability~\cite{Potter_2022}, reducing future transaction flow and harming long-term adoption.%Consequently, while the immediate financial losses of liquidity exhaustion attacks are small, their reputational and ecosystem-level effects may be substantial and persistent.

\begin{findingbox}
\textbf{Main Insights.} A byzantine adversary can degrade availability across all evaluated intent-based bridges, causing dozens of failed intents per attack window and affecting high-value transfers due to the heavy-tailed distribution of intent values. Although direct financial losses to solvers and protocols are modest, even short-lived disruptions pose a broader systemic risk by harming user experience and potentially generating lasting reputational and ecosystem-level consequences.
\end{findingbox}

\section{Targeted Liquidity Exhaustion Attacks}
\label{sec: targeted_attack}

%SHAP values demonstrate that the total solver liquidity that we need to drain is the main cause of cost for the attack. Therefore, here we explore some solver strategies that can be exploited to decrease the attack cost.

The baseline strategy introduced in Section~\ref{sec: baseline_strategy} relies on solver liquidity statistics and exploits naturally occurring liquidity dips. Although effective, this approach implicitly assumes that all solvers compete uniformly across all intents, and therefore all liquidity needs to be temporarily drained.

\subsection{Solver Participation Patterns}

In practice, solver participation is highly structured and depends on intent-specific attributes. We therefore introduce a \emph{targeted attack strategy} that exploits these participation patterns. By targeting specific intent characteristics where not all solvers compete in the auction, the attacker only needs to consider the liquidity of the competing solvers, which substantially reduces the cost of the attack. These patterns focus on three dimensions:

\begin{itemize}\setlength\itemsep{-0.1em}
    \item \textbf{intent value thresholds}, as solvers selectively compete on specific value ranges of intents.
    \item \textbf{solver activity windows}, because some solvers are not continuously active and temporarily halt bidding.
    \item \textbf{token specialization}, whereby solvers compete only on a subset of tokens.% to avoid maintaining liquidity in multiple tokens
\end{itemize}

% probably give a new name to this section....
To concretely evaluate the targeted strategy, we focus on specific intent classes described below.

\begin{figure}[h]
    \centering
    \includegraphics[width=1.1\linewidth]{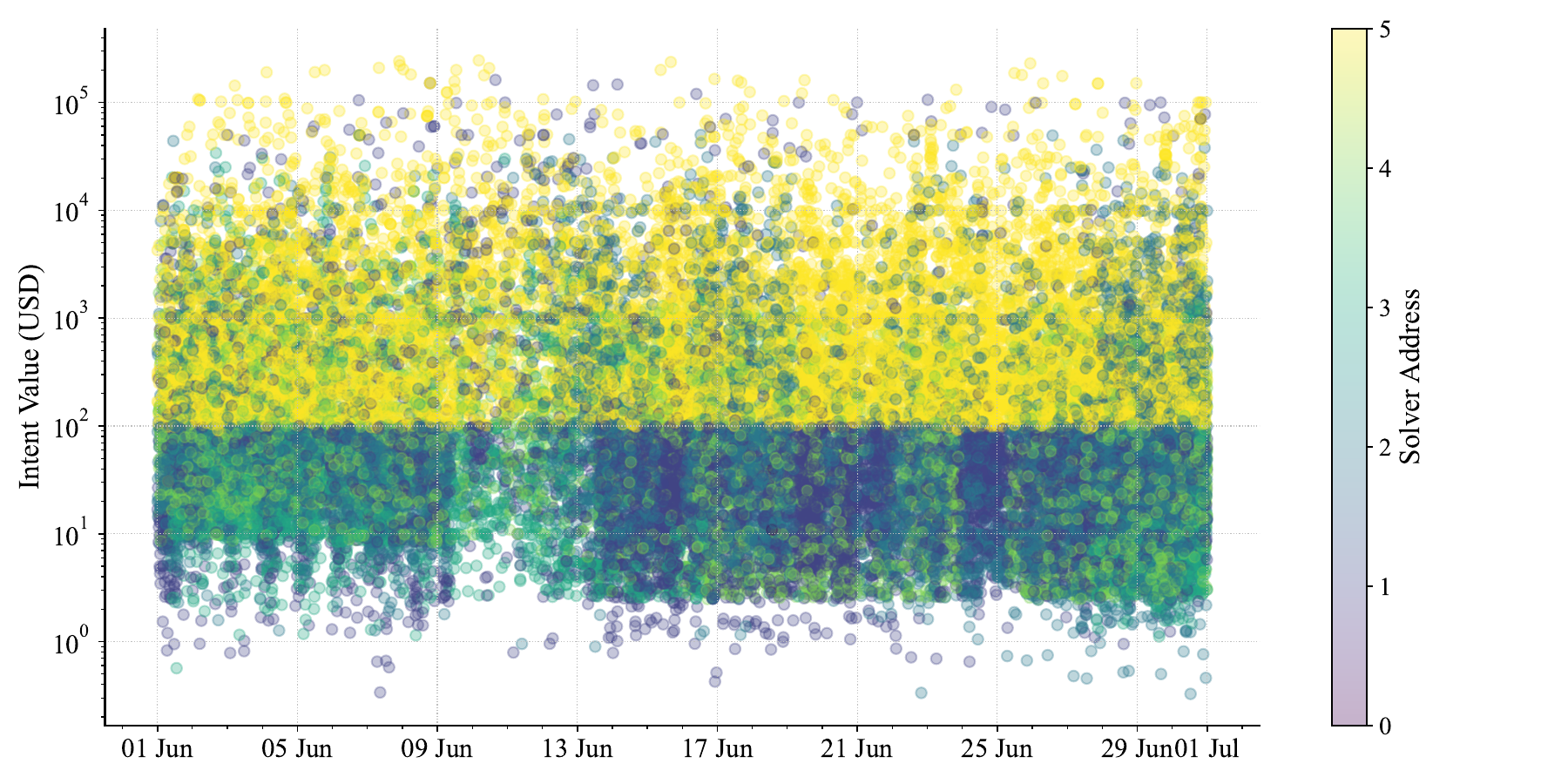}
    \caption{Scatter plot of auction bids for the top 5 solvers of Mayan Swift in June 2025. Each dot represents a bid colored by the solver address.}
    \label{fig: solver_strategy}
\end{figure}

\featheader{Mayan Swift}
We consider transactions of USDC in Mayan Swift with an intent value lower than \$100. This class is particularly suitable for studying targeted attacks for three reasons. First, USDC is one of the most commonly used destination tokens on Mayan Swift, ensuring sufficient data for statistical analysis. Second, intents below \$100 exhibit reduced solver competition compared to higher-value transfers (cf. Figure~\ref{fig: solver_strategy}). Third, while individual intent values are small, these intents occur frequently (cf. Table~\ref{tab: protocols-summary}), making them relevant from an availability perspective even if profit potential is limited. From our data analysis, we found the following patterns: (1) \AddrHrefEthereum{0xdfd122610a14ac12d934898c02dbec1f72708116} does not bid on intents lower than \$100. Its total liquidity sits between $\approx\$500,000$ and $\approx\$1,000,000$ in the interval under analysis; (2) solvers \AddrHrefEthereum{0xcbb0cb4492afbcd9963441cc6aea50f35807ff96} and \AddrHrefEthereum{0x38bf020e39e5a3ef1519c1283f6cac8a6b5851ff} only made bids and fulfillments at the beginning of the interval of analysis and were kept deactivated until the end. We also noted that solver \AddrHrefEthereum{0x7c825c6e7e4e1f618ca67e4943cdb41ca00b7f6b} had occasional periods of inactivity. However, since these periods were not predictable, an attacker could not leverage that information prior to performing an attack. With the information above, we exclude the total liquidity of the solvers where patterns can be observed, resulting in a decrease in the total liquidity that needs to be temporarily drained by $\mathbb{A}$. Recalling the liquidity exhaustion parameter $\alpha$ introduced in Equation~\ref{eq: cost_induction}, under the patterns above for Mayan Swift and $k = 2$, the median value of $\alpha$ is $0.075$ -- i.e., the attacker is only required to exhaust $\approx 7.5\%$ (\$57k) of the total liquidity available).

\featheader{deBridge}
Similarly to Mayan Swift, solver participation in deBridge exhibits strong intent value-based trends. We focus on USDC intents with token values above \$500, which reduces the number of competing solvers. USDC is the most frequently used token in deBridge, ensuring sufficient data for analysis. Solvers \AddrHrefEthereum{0xc4eb49ea01578cb9b1c68ad27f457dbfa0bfbd97} and \AddrHrefEthereum{0x78b0f42536aeee037deedbb968ffb23cc2c0082e} exclusively bid on and fulfill USDC intents with values below \$500 and do not participate in higher-value transfers. In contrast, the solver \AddrHrefEthereum{0x555ce236c0220695b68341bc48c68d52210cc35b} remains consistently active throughout the interval and fulfills intents throughout the entire value range. Under these parameters and $k = 1$ ($k = 2$ did not yield results), the median value of $\alpha$ is $0.129$ -- i.e., the attacker is only required to exhaust $\approx 12.9\%$ (\$41k) of the total liquidity available.

\featheader{Across} Unlike other protocols, Across does not exhibit clear solver participation patterns. Although some solvers are active only intermittently, their behavior is insufficiently predictable to enable reliable targeting strategies.

\subsection{Targeted Attack Model}

We extend the model in Section~\ref{sec: baseline_strategy} by conditioning solver liquidity on intent characteristics. Let $\mathcal{S}$ denote the set of all solvers, and $\mathcal{S}(c) \subseteq \mathcal{S}$ be the subset of solvers that historically compete for intents of class $c$, where $c$ encodes the intent value threshold, solver activity windows, and tokens used.

Let $\mathcal{S}(c)$ denote solvers that historically compete for intent class $c$. The effective competing liquidity is:
\begin{equation}
L_{\mathrm{eff}}(c,t) = \sum_{s \in \mathcal{S}(c)} L_s(t).
\end{equation}

Similarly to the baseline strategy, $\mathbb{A}$ triggers a targeted attack whenever:
\begin{equation}
L_{\mathrm{eff}}(c,t) < \tilde{L}_{\mathrm{eff}}(c) - k \cdot \sigma_{\mathrm{eff}}(c),
\end{equation}

\subsection{Simulation Results}

Table~\ref{tab: results-usdc} summarizes the results of the targeted attack strategy under the same baseline parameters ($k=1$, 1000s attack window). Overall, targeting solver participation substantially reduces attack costs, but does not uniformly translate into profitable attacks across protocols.

\featheader{Mayan Swift} The targeted strategy remains largely unprofitable. In all configurations, the mean net profit ranges from approximately -\$31 to -\$15, with profitability probabilities below 25\%. Even when transaction volume is artificially increased (e.g., volume multiplier of 5.61), expected profit remains negative, although the upper tail becomes marginally positive ($p90 \approx \$17$). This outcome is primarily driven by the low economic value of the target intent class (transfers below \$100). Although targeting successfully reduces the liquidity that must be drained, the attainable solver revenue is inherently capped by the small transaction sizes. As a result, cost reductions alone are insufficient to make the attack economically viable. These findings suggest that Mayan Swift is structurally resilient to targeted liquidity exhaustion attacks when low-value intents dominate the attack surface.

\featheader{deBridge} In contrast, deBridge exhibits strong susceptibility to targeted attacks. Under real protocol parameters, the attack yields a mean net profit of up to \$210 with near-certain long-term profitability (96 -- 100\%). Even when transaction volume is reduced (volume multiplier 0.178), the attack remains profitable (\$20.94 mean profit with 67\% success probability). The primary driver of the success is the high value of captured intents: the attacker fulfills on average \$86k per window, over two orders of magnitude higher than in Mayan Swift. Consequently, even moderate solver margins generate sufficient revenue to offset attack costs. Profitability persists across most stress-test configurations, including higher protocol fees and alternative solver profitability assumptions. Only when solver profitability is drastically reduced (0.018\%) does the attack become borderline unprofitable, reinforcing the importance of solver margins as a first-order security parameter.

\begin{findingbox}
\textbf{Main Insights.} The targeted strategy is radically constrained by the economic value of the intents captured. Reducing the competing liquidity is necessary but not sufficient for profitability, as high transaction values and solver margins remain decisive factors. However, these results reinforce our earlier insight that protocol security depends heavily on external usage patterns rather than protocol mechanics alone. Protocols processing large, high-margin transfers are inherently more exposed to economically motivated attackers.%, even when solver participation is sparse.
\end{findingbox}

\begin{comment}
The results 

For Across and  protocols, this targeted approach consistently outperforms the baseline strategy.

However, expected profit remains near zero: intent classes with sparse competition also have low economic value.

Targeted attacks reveal a fundamental asymmetry: the cheapest intents to attack are also the least profitable. While attackers can substantially reduce capital requirements, this rarely translates into economically viable attacks. Nevertheless, these strategies remain relevant from an availability perspective, enabling low-cost and reliable denial-of-service attacks against specific intent classes.

\begin{findingbox}
\textbf{Main Insights.} xxx
\end{findingbox}
\end{comment}

\input{tables/main-body-results-dos-usdc}

\subsection{Massive Attack Cost Reduction}

\begin{figure}[ht]
    \centering
    \includegraphics[width=\linewidth]{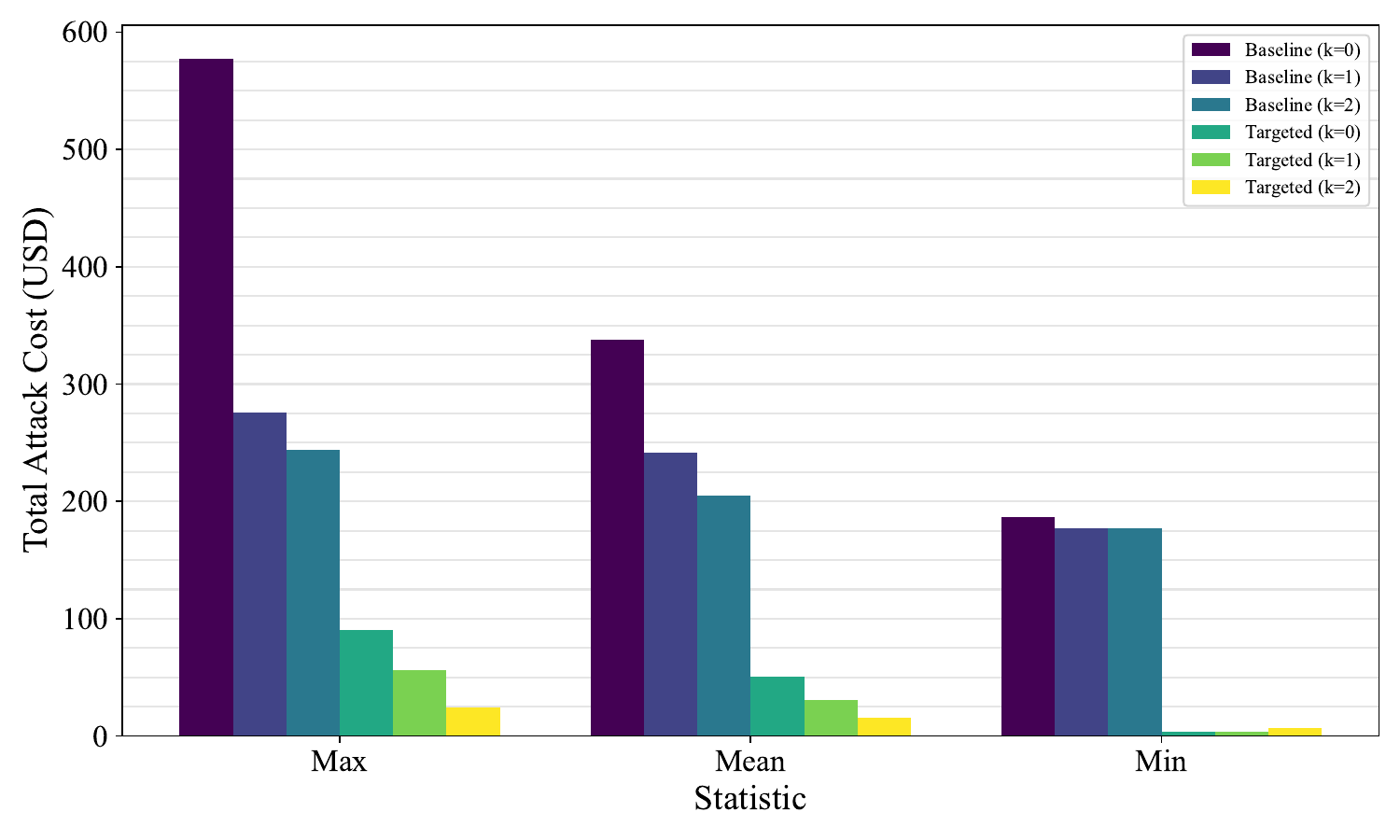}
    \caption{Cost comparison between baseline (rational and byzantine) and targeted strategies for attacks to Mayan Swift.}
    \label{fig: cost_reduction_targeted}
\end{figure}

Figure~\ref{fig: cost_reduction_targeted} compares the min, max and median value of the total cost of an attack in the baseline and targeted strategies for $k \in \{0,1,2\}$ in Mayan Swift. This optimized strategy massively reduces the attack costs. As an example, for $k = 2$, the minimum cost for performing the attack against Mayan Swift decreased from $\$175$ to $\$7$, while the maximum reduced from $\$247$ to $\$25$. In deBridge, for $k = 0$ and $k = 1$, the maximum dropped 90.41\% (\$1507.32 to \$144.56) and 77.19\% (\$158.31 to \$57.17), respectively. The mean reduced 77.19\% (\$266.38 to \$60.76) and 80.74\% (\$88.77 to \$17.10), respectively.

\begin{comment}
\begin{figure*}[t]
    \centering
    \includegraphics[width=\linewidth]{figures/comparison-baseline-vs-targeted.png}
    \caption{Cost comparison between baseline (rational and byzantine) and targeted strategies for attacks to Mayan Swift.}
    \label{fig: cost_reduction_targeted_v2}
\end{figure*}
\end{comment}

\begin{findingbox}
\textbf{Main Insights.} Regardless of the motivation of the attacker (i.e., rational or byzantine), the targeted attack strategy is able to reduce the minimum, median, and maximum cost of performing an attack on Mayan Swift by 97.18\%, 90.24\%, and 82.7\%, respectively.
\end{findingbox}

% mean (205-20)/205x100 = 90.24%
% min (177-5)/177x100 = 97.18%
% max (145-25)/145x100 = 82.7%

\section{Defense Strategies}
\label{sec: defense_strategies}

Liquidity exhaustion attacks exploit a core property of intent-based bridges: solvers front liquidity and wait for asynchronous settlement. As a result, defenses must preserve solver liquidity and protocol availability without sacrificing the low-latency that motivated intent-based protocols. Below, we discuss several defense strategies based on our results.

\featheader{Liquidity-Aware Rate Limiting}
A first option is to rate-limit intent fulfillment based on available solver liquidity. Although liquidity-aware throttling can prevent complete liquidity shortages, it does not eliminate the possibility of the attack, as adversaries can still exploit naturally occurring liquidity dips. Such mechanisms reduce throughput when demand is high, exchanging availability failures for a slower service.

\featheader{Accelerated or Conditional Refunds}
Reducing refund times (by enabling partial or conditional refunds) directly shrinks the attack window, and thus lowers attacker profitability. However, changing the refund time window is a difficult security challenge. Refunds are dependent on the time needed to validate the cross-chain transaction and ensure the finality of the transaction on the source blockchain.

\featheader{Automated Solver Rebalancing}
Our measurements indicate that solver rebalancing and liquidity injections are largely manual and weakly correlated with low-liquidity conditions (cf. Figure~\ref{fig: liquidity_injections_native_mayan}). Automating rebalancing, either across chains or from external liquidity sources, can substantially reduce the duration of exploitable liquidity dips. Although small reductions in downtime have a limited impact, fully automated responses can significantly reduce the probability of success of the attack by shortening the attack window.

\featheader{Dynamic Fee Adjustment}
Similar to congestion pricing~\cite{eip_1559} or lending protocols interest rate calculation~\cite{aave}, intents that consume large amounts of liquidity should be more expensive, and changing protocol fees based on the total liquidity available. At the same time, from our measurements, very low solver profitability turns the attack practically infeasible. The sweat spot between low solver profitability, but still attracts new entities to act as solvers, needs to be researched.

\featheader{Solver Diversity and Liquidity Scale}
Solver concentration amplifies the impact of liquidity exhaustion attacks. Increasing solver diversity and, most importantly, the total available liquidity, raises the cost of performing the attack.

\featheader{Discussion}
No mitigation fully prevents liquidity exhaustion attacks without sacrificing some benefits of intent-based designs. Instead, effective defense requires both protocol-level safeguards and solver-side operational practices. Our results suggest that maintaining low solver profitability, reducing settlement latency, and increasing total liquidity are among the most effective strategies against rational attackers. 

%Importantly, liquidity exhaustion is detectable. Attacks rely on observable signals such as rapid liquidity depletion, abnormal intent volume, or repeated large-value intents. Protocols and solvers can therefore deploy real-time monitoring and risk assessment systems to detect attack conditions early. Our simulation framework can be repurposed to estimate attack profitability under current conditions and trigger automated responses -- such as throttling, rebalancing, or fee adjustments -- when predefined risk thresholds are exceeded.

\section{Discussion}
\label{sec:discussion}

In this section, we discuss our findings and some limitations.

%Because solvers front capital while awaiting asynchronous settlement, temporary liquidity shortages can translate directly into availability failures. These shortages may arise from demand fluctuations and are not consistently mitigated by automated solver rebalancing in current deployments.

In this paper, we show that risk can emerge even in the absence of software vulnerabilities or adversarial manipulation of protocol logic. We also show that, in some cases, internal parameters alone are insufficient to safeguard solvers and protocols. In these instances, protocol security depends heavily on external factors such as transaction volume, intent value distributions, and temporal demand patterns. This external dependency challenges the notion that intent-based protocols can be secured solely through internal parameter tuning.

Even when an attack is not profitable, byzantine adversaries can degrade availability by suppressing solver participation, and cause substantial indirect effects. Bridges increasingly compete for order flow routed by aggregators that prioritize reliability. Short-lived disruptions can cause reputational damage and persistent traffic loss. Risk is further amplified by the heavy-tailed distribution of intent values. Although most failed intents are small (cf. Section~\ref{tab: byzantine_p90}), a few high-value transfers dominate user impact. Security outcomes are thus shaped more by tail events than by average behavior.

Solver profitability is very influential in the success of liquidity exhaustion attacks. Low margins discourage economically rational attackers but may also reduce solver participation, revealing an inherent trade-off between economic viability of a project and resilience to attacks. Our findings suggest that mitigating liquidity exhaustion requires both protocol-level safeguards and solver-side operational practices. Mechanisms inspired by usual defenses, such as rate limiting, circuit breakers, and risk-adjusted pricing, could limit exposure during liquidity stress. Complementary accounting-based approaches, including solver balance invariants, may further constrain the conditions under which liquidity shortages propagate into protocol-wide failures. Together, these strategies point toward liquidity-aware architectures that treat capital constraints as an explicit component of system security.

\featheader{Limitations}
\label{subsec:limitations}
Liquidity exhaustion attacks require multiple conditions to align, including sufficient transaction volume and limited liquidity rebalancing, and may not persist under rapidly changing market conditions. Our replay framework models solvers as black boxes and \textit{does not capture adaptive responses such as surge pricing, throttling, or rapid capital injection}. However, we focus on short attack windows, under 1{,}000 seconds, based on the solver refund time in Table~\ref{tab: protocols-summary}. Although Across, theoretically, has a much longer settlement period (2 hours), we believe sustained liquidity shortages over such intervals without intervention are unrealistic.

%% WIP - TO ADD A SENTENCE WITH THE THOUGHT BELOW
% A question the reviewers may have is: why didnt you try to exhaust just a portion of the liquidity instead of always the entire liquidity? You could model the auction to understand if the attacker would win a certain auction or not.

%ANSWER: Throughout the experiments, assume to drain the total liquidity (or exhaust the liquidity of any competing solvers, to guarantee that the attacker wins the auction). Effectively, an alternative is to model the auction and make auction winning predictions. The hypothesis behind is that if I drain more liquidity, I will have more chances of winning the auction. Draning less liquidity will yield less chances of winning. And we would find the breaking point. We tried this, however, no auction validated this hypothesis, meaning that there was no correlation between winning auction and draining more or less liquidity. If we know the way to win all auctions, then we can just do that, we don't need to spend funds exhausting the liquidity...

A broader challenge is the limited observability of solver behavior: liquidity policies, rebalancing strategies, and risk thresholds are not public. By relying exclusively on on-chain data, our framework mirrors what real adversaries see. While this abstraction prevents precise prediction, our data-driven patterns show that liquidity exhaustion is a credible risk for protocol designers and operators.

\section{Related Work}
\label{sec: related_work}

Prior work has extensively studied blockchain interoperability, including cross-chain communication~\cite{zamyatin2021sok}, bridge architectures~\cite{belchior2023you}, trust assumptions~\cite{belchior2021survey}, and known vulnerability classes~\cite{augusto_sok_2024}. We build on these foundations by examining an underexplored risk in intent-based protocols: the temporary exhaustion of solver liquidity. Related liquidity-driven attacks have previously been identified. In Hash Time Locked Contracts (HTLCs), adversaries can immobilize counterparties’ funds through \emph{Griefing} or \emph{Sore Loser Attacks}~\cite{herlihy2018atomic,xue2021hedging,mazumdar2022towards}, while payment channel networks have been shown to suffer channel depletion attacks that restrict routing capacity~\cite{rohrer2019discharged}.

Intent-based systems shift execution off-chain into competitive markets where solvers submit bids to satisfy user intents. Chitra et al.~\cite{chitra2024analysis} provide one of the first analyses of intent-based market mechanisms, and Canidio and Henneke~\cite{Canidio_Henneke_2024} study the economics of trade-intent auctions. Additional work explores decentralized solver architectures and emerging standardization efforts~\cite{economics_2023}, but largely abstracts away from solver balance-sheet constraints and their interaction with delayed settlement. Yuminaga et al.~\cite{yuminaga2025execution} show that solver-based DEXes can improve execution welfare relative to AMM-based designs. Finally, research on maximal extractable value (MEV) primarily characterizes adversarial profit opportunities arising from transaction ordering~\cite{qin2022quantifying,10.1145/3658644.3690259}. In contrast, we consider rational adversaries that exploit volatile solver liquidity and settlement delays rather than transaction ordering.

%\paragraph{Our Work}
To our knowledge, this paper is the first to systematically study liquidity exhaustion as an attack vector in intent-based cross-chain protocols. We show how delayed settlement and finite solver capital can be leveraged to induce protocol-level availability failures, even without network congestion or smart contract vulnerabilities.

\section{Conclusion}
\label{sec: conclusion}
Intent-based cross-chain bridges mark a shift in interoperability design, prioritizing speed and user experience over synchronous verification. We show that this shift introduces vulnerabilities rooted in solver liquidity.%: because solvers front capital and rely on delayed refunds, their temporary exhaustion can degrade or deny protocol availability.
Through empirical analysis of millions of cross-chain intents and replay-based simulations, we propose \emph{liquidity exhaustion attacks} and demonstrate their feasibility in current deployments. Exploiting natural liquidity fluctuations can disrupt availability and, under some circumstances, generate profit. Vulnerability varies significantly across protocols, with low solver margins discouraging rational attackers, while higher profitability and liquidity concentration increase risk. Surprisingly, security often depends more on external factors, such as traffic patterns and solver behavior, than on protocol parameters alone. Mitigating these risks requires both protocol-level safeguards and solver-side practices, including automated rebalancing, faster refunds, and liquidity-aware rate limiting. More broadly, our findings position liquidity exhaustion as a fundamental availability risk that should be treated as a first-class security consideration in the design of future intent-based systems.

%% file: tables/protocols-summary.tex
\setlength{\tabcolsep}{5.5pt} % Default value: 6pt

\begin{table}[ht]
\scriptsize
\begin{tabular}{@{}lccc@{}}
\toprule
 & \multicolumn{1}{c}{Mayan Swift} & \multicolumn{1}{c}{Across} & \multicolumn{1}{c}{deBridge} \\ \midrule
Solver Selection Algorithm & English Auction & FCFS & FCFS \\
Bridge Protocol & Wormhole & Across & deBridge \\ 
\# of Unique Solvers$\dagger$ & 14 & 60 & 9 \\
\% of Fulfills by Top Solver & 24 & 19 & 94 \\
\# of Fulfills (Jun - Nov 2025) & 895,206 & 1,619,496 & 937,688\\
\hdashline
Median Intent Fulfillment Time (s) & 28.0 & 8.0 & 20.0 \\
Median Solver Refund Time (s) & 1281 & 7200* & 989 \\
Median Solver Profitability (\%) & 0.381 & 0.018 & 1.129 \\
Median User Paid Protocol Fees (\%) & 0.029 & 0.027 & 1.305 \\
\hdashline
Median Intent Value (\$) & 74.395 & 73.060 & 260.411 \\
Median Total Solver Liquidity (\$) & 600,000 & 8,900,000 & 514,000 \\
Total Value Transacted (Billion \$) & $0.852$ & $3.61$ & $4.78$ \\
\bottomrule
\end{tabular}
\begin{tablenotes}
            \footnotesize
            \item *theoretical value based on Across's documentation
            \item $\dagger$ a complete list can be found in Table~\ref{tab: solver_addreses} in Appendix~\ref{appendix: addresses}

\end{tablenotes}
\caption{Stats for the bridges analyzed and the stats are presented for fulfills on Ethereum}
\label{tab: protocols-summary}
\end{table}

%% file: tables/k-sensitivity.tex
\setlength{\tabcolsep}{2pt} % Default value: 6pt

\begin{table}[ht]
\centering
\scriptsize
\rowcolors{2}{white}{white}
\begin{tabular}{lrrrrr}
\toprule
Bridge & $k$ & Mean Net Profit (\$) & Volume Fulfilled (\$) & $\Pr[\text{Profit}]$ & Reliable Attack? \\
\midrule
\rowcolor{gray!8}
Mayan & 0 & -295.66 $\pm$ 103.60 & 34,989 & 1.1\% & No \\
\rowcolor{gray!8}
Mayan & 1 & -195.20 $\pm$ 154.37 & 37,106 & 1.3\% & No \\
\rowcolor{gray!8}
Mayan & 2 & -178.68 $\pm$ 25.96 & 20,873 & 0.0\% & No \\
\rowcolor{gray!8}
Mayan & 3 & -185.95 $\pm$ 3.75 & 24,086 & 0.0\% & No \\
Across & 0 & -2828.81 $\pm$ 644.96 & 218,253 & 0.0\% & No \\
Across & 1 & -2828.81 $\pm$ 644.96 & 218,253 & 0.0\% & No \\
Across & 2 & -2719.64 $\pm$ 559.76 & 219,901 & 0.0\% & No \\
Across & 3 & 0 & 0 & 0 & No \\
\rowcolor{gray!8}
Debridge & 0 & \cellcolor{green!15}199.77 $\pm$ 1316.75 & 138,537 & 45.2\% & No \\
\rowcolor{gray!8}
Debridge & 1 & \cellcolor{green!15}286.14 $\pm$ 1075.66 & 107,353 & \cellcolor{blue!15}80.5\% & \cellcolor{blue!15}Yes \\
\rowcolor{gray!8}
Debridge & 2 & \cellcolor{green!15}203.06 $\pm$ 244.56 & 92,623 & \cellcolor{blue!15}100.0\% & \cellcolor{blue!15}Yes \\
\rowcolor{gray!8}
Debridge & 3 & 0 & 0 & 0 & No \\
\bottomrule
\end{tabular}
\caption{Sensitivity of liquidity exhaustion attacks to the attack parameter $k$ for attack window = 1000s, and real protocol parameters. Cell shading highlights profitable outcomes and high attack reliability.}
\label{tab: k_sensitivity_1000}
\end{table}

\begin{table}[ht]
\centering
\scriptsize
\rowcolors{2}{white}{white}
\begin{tabular}{lrrrrr}
\toprule
Bridge & $k$ & Mean Net Profit (\$) & Volume Fulfilled (\$) & $\Pr[\text{Profit}]$ & Reliable Attack? \\
\midrule
\rowcolor{gray!8}
Mayan & 0 & -325.59 $\pm$ 74.55 & 10,589 & 0.2\% & No \\
\rowcolor{gray!8}
Mayan & 1 & -229.91 $\pm$ 29.67 & 10,198 & 0.0\% & No \\
\rowcolor{gray!8}
Mayan & 2 & -195.97 $\pm$ 20.75 & 10,199 & 0.0\% & No \\
\rowcolor{gray!8}
Mayan & 3 & -198.48 $\pm$ 8.49 & 17,449 & 0.0\% & No \\
Across & 0 & -2882.98 $\pm$ 628.84 & 63,922 & 0.0\% & No \\
Across & 1 & -2882.98 $\pm$ 628.84 & 63,922 & 0.0\% & No \\
Across & 2 & -2773.42 $\pm$ 542.51 & 63,414 & 0.0\% & No \\
Across & 3 & 0 & 0 & 0 & No \\
\rowcolor{gray!8}
Debridge & 0 & -132.26 $\pm$ 505.54 & 41,633 & 12.1\% & No \\
\rowcolor{gray!8}
Debridge & 1 & \cellcolor{green!15}84.41 $\pm$ 986.55 & 42,684 & 26.2\% & No \\
\rowcolor{gray!8}
Debridge & 2 & -13.91 $\pm$ 36.07 & 21,051 & 20.0\% & No \\
\rowcolor{gray!8}
Debridge & 3 & 0 & 0 & 0 & No \\
\bottomrule
\end{tabular}
\caption{Sensitivity of liquidity exhaustion attacks to the attack aggressiveness parameter $k$ for attack window = 300s. Cell shading highlights profitable outcomes and high attack reliability.}
\label{tab: k_sensitivity_300}
\end{table}

%% file: tables/main-body-results-dos.tex
\setlength{\tabcolsep}{4pt} % Default value: 6pt

\begin{table*}[ht]
\scriptsize
\rowcolors{2}{gray!10}{white}
\centering
\begin{tabular}{lccccccrrrrr}
\toprule
& & \multicolumn{4}{c}{\textbf{Simulated Values}} & \multicolumn{5}{c}{\textbf{Simulation Results}}\\ 
\cmidrule(rl){3-6} \cmidrule(l){7-11}
Bridge & Src → Dst & S Profit. & Prot. Fee & Max Tx Value & Vol Mul & N. Attacks & N. Fulfillments & Volume Fulfilled (\$) & Net Profit (\$)  & $Pr[\text{Profit}]$ \\
\midrule
mayan & solana → ethereum & Real & 0.027 & 10000 & 1 & 227 & 23.84 ± 7.57 & 37105.94 ± 48188.45 & \cellcolor{red!0}{-178.56 (p90: -145.72)} & 1.3\% \\

mayan & solana → ethereum & 0.018 & Real & 10000 & 1 & 227 & 23.84 ± 7.57 & 37105.94 ± 48188.45 & \cellcolor{red!0}{-235.15 (p90: -202.77)} & 0.0\% \\

mayan & solana → ethereum & Real & Real & 10000 & 5.61 & 227 & 133.73 ± 42.48 & 208164.33 ± 270337.22 & \cellcolor{green!5}{19.74 (p90: 191.13)} & 23.8\% \\

\rowcolor{gray!30}
mayan & solana → ethereum & Real & Real & 10000 & 1 & 227 & 23.84 ± 7.57 & 37105.94 ± 48188.45 & \cellcolor{red!0}{-195.2 (p90: -161.37)} & 1.3\% \\

mayan & solana → ethereum & Real & Real & 100000 & 1 & 227 & 23.84 ± 7.57 & 37105.94 ± 48188.45 & \cellcolor{red!0}{-201.45 (p90: -165.57)} & 1.3\% \\

mayan & solana → ethereum & 1.129 & Real & 10000 & 1 & 227 & 23.84 ± 7.57 & 37105.94 ± 48188.45 & \cellcolor{green!48}{177.1 (p90: 852.08)} & 48.9\% \\

mayan & solana → ethereum & Real & 1.3 & 10000 & 1 & 227 & 23.84 ± 7.57 & 37105.94 ± 48188.45 & \cellcolor{red!12}{-10795.83 (p90: -9500.89)} & 0.0\% \\

mayan & solana → ethereum & Real & Real & 10000 & 4.237 & 227 & 101.0 ± 32.09 & 157217.87 ± 204174.47 & \cellcolor{red!0}{-44.27 (p90: 84.48)} & 17.2\% \\

debridge & solana → ethereum & 0.381 & Real & 10000 & 1 & 210 & 27.55 ± 8.58 & 107352.6 ± 374202.64 & \cellcolor{green!87}{320.24 (p90: 625.91)} & \cellcolor{blue!15}72.4\% \\

debridge & solana → ethereum & Real & Real & 10000 & 0.755 & 210 & 20.8 ± 6.48 & 81051.21 ± 282522.99 & \cellcolor{green!53}{194.29 (p90: 443.7)} & \cellcolor{blue!15}67.1\% \\

\rowcolor{gray!30}
debridge & solana → ethereum & Real & Real & 10000 & 1 & 210 & 27.55 ± 8.58 & 107352.6 ± 374202.64 & \cellcolor{green!78}{286.14 (p90: 617.46)} & \cellcolor{blue!15}80.5\% \\

debridge & solana → ethereum & Real & 0.029 & 10000 & 1 & 210 & 27.55 ± 8.58 & 107352.6 ± 374202.64 & \cellcolor{green!77}{281.55 (p90: 615.17)} & \cellcolor{blue!15}77.6\% \\

debridge & solana → ethereum & Real & Real & 100000 & 1 & 210 & 27.55 ± 8.58 & 107352.6 ± 374202.64 & \cellcolor{green!100}{364.77 (p90: 699.83)} & \cellcolor{blue!15}99.0\% \\

debridge & solana → ethereum & 0.018 & Real & 10000 & 1 & 210 & 27.55 ± 8.58 & 107352.6 ± 374202.64 & \cellcolor{red!0}{-69.45 (p90: -35.11)} & 1.9\% \\

debridge & solana → ethereum & Real & Real & 50000 & 1 & 210 & 27.55 ± 8.58 & 107352.6 ± 374202.64 & \cellcolor{green!97}{356.04 (p90: 691.65)} & \cellcolor{blue!15}99.0\% \\

debridge & solana → ethereum & Real & 0.027 & 10000 & 1 & 210 & 27.55 ± 8.58 & 107352.6 ± 374202.64 & \cellcolor{green!78}{287.99 (p90: 621.89)} & \cellcolor{blue!15}80.5\% \\

debridge & solana → ethereum & Real & Real & 10000 & 0.178 & 210 & 4.9 ± 1.53 & 19108.76 ± 66608.07 & \cellcolor{red!0}{-22.04 (p90: 41.51)} & 18.1\% \\

\rowcolor{gray!30}
across & base → ethereum & Real & Real & 10000 & 1 & 2440 & 43.45 ± 11.11 & 218253.35 ± 390200.38 & \cellcolor{red!3}{-2828.81 (p90: -2076.13)} & 0.0\% \\

across & base → ethereum & Real & Real & 10000 & 1.323 & 2440 & 57.48 ± 14.7 & 288749.18 ± 516235.1 & \cellcolor{red!3}{-2804.57 (p90: -2041.51)} & 0.0\% \\

across & base → ethereum & 1.129 & Real & 10000 & 1 & 2440 & 43.45 ± 11.11 & 218253.35 ± 390200.38 & \cellcolor{red!0}{-439.79 (p90: 2705.82)} & 23.4\% \\

across & base → ethereum & Real & 1.3 & 10000 & 1 & 2440 & 43.45 ± 11.11 & 218253.35 ± 390200.38 & \cellcolor{red!100}{-87464.8 (p90: -68160.2)} & 0.0\% \\

across & base → ethereum & 0.381 & Real & 10000 & 1 & 2440 & 43.45 ± 11.11 & 218253.35 ± 390200.38 & \cellcolor{red!2}{-2072.32 (p90: -724.92)} & 6.1\% \\

across & base → ethereum & Real & Real & 10000 & 0.236 & 2440 & 10.25 ± 2.62 & 51507.79 ± 92087.29 & \cellcolor{red!3}{-2886.16 (p90: -2150.51)} & 0.0\% \\

across & base → ethereum & Real & Real & 100000 & 1 & 2440 & 43.45 ± 11.11 & 218253.35 ± 390200.38 & \cellcolor{red!3}{-3157.64 (p90: -1299.29)} & 0.0\% \\

across & base → ethereum & Real & 0.029 & 10000 & 1 & 2440 & 43.45 ± 11.11 & 218253.35 ± 390200.38 & \cellcolor{red!2}{-1877.75 (p90: -1430.73)} & 0.0\% \\

\bottomrule
\end{tabular}
\begin{tablenotes}
            \footnotesize
            \item \textit{Note:} The table includes both observed and counterfactual parameter configurations. For each protocol, results are reported using its real on-chain parameters, as well as simulations where individual economic/usage variables -- solver profitability (S Profit.), protocol fee (Prot. Fee), maximum intent value submitted (Max Tx Value), and the volume multiplier (Vol Mul) -- are replaced with values observed in other protocols. These substitutions allow us to isolate the effect of specific parameters on attack outcomes while keeping liquidity dynamics and transaction distributions fixed. Counterfactual rows should be interpreted as stress tests rather than expected real-world behavior. Gray table lines mark simulations with the real protocol parameters.
            \emph{Real solver profitability values:} Mayan 0.381\%, DeBridge 1.129\%, Across 0.018\%.
            \emph{Real protocol fee values:} Mayan 0.029\%, DeBridge 1.3\%, Across 0.027\%.

\end{tablenotes}

\caption{Simulation Results for $k$ = 1 and AttWindow = 1000 with Color-Coded Net Profit}
\label{tab: results}
\end{table*}

%% file: tables/results-byzantine.tex
\setlength{\tabcolsep}{4pt} % Default value: 6pt

\begin{table*}[ht]
\scriptsize
\rowcolors{2}{gray!10}{white}
\centering
\begin{tabular}{lrrrrrrr}
\toprule
Bridge & Att. Window (s) & Total Cost (\$) & Failed Intent Value (\$) & Failed Intent Value Std (\$) & Failed Intents & Missed Solver Profit (\$) & Missed Protocol Fees (\$) \\
\midrule
across & 200 & 2891.80 / 3658.51 & 116.82 / 644.94 & 1592.57 / 26809.87 & 8.0 / 13.0 & 3.95 / 27.66 & 12.27 / 55.75 \\
across & 600 & 2886.00 / 3654.33 & 107.60 / 347.08 & 5083.01 / 37535.04 & 25.0 / 36.0 & 16.74 / 99.60 & 46.01 / 171.71 \\
across & 1000 & 2886.00 / 3654.33 & 109.68 / 295.54 & 8178.19 / 43597.76 & 42.5 / 58.0 & 34.89 / 161.67 & 83.40 / 285.75 \\
debridge & 200 & 239.45 / 397.01 & 197.83 / 1216.04 & 1235.78 / 20452.17 & 5.0 / 9.0 & 27.37 / 200.99 & 19.75 / 41.17 \\
debridge & 600 & 239.45 / 395.56 & 173.21 / 515.42 & 3306.75 / 28230.67 & 16.0 / 25.0 & 113.28 / 561.47 & 61.03 / 116.38 \\
debridge & 1000 & 239.45 / 395.56 & 174.52 / 450.83 & 4992.74 / 34177.23 & 27.0 / 40.0 & 212.10 / 978.97 & 104.75 / 187.98 \\
mayan & 200 & 339.89 / 444.50 & 84.03 / 546.91 & 392.51 / 5510.47 & 5.0 / 8.0 & 3.12 / 15.36 & 0.36 / 4.76 \\
mayan & 600 & 337.30 / 440.16 & 79.27 / 285.74 & 1308.09 / 10718.09 & 13.0 / 22.0 & 11.98 / 57.52 & 2.26 / 15.84 \\
mayan & 1000 & 337.35 / 440.12 & 74.28 / 223.65 & 1944.28 / 13025.88 & 23.0 / 35.0 & 21.76 / 89.05 & 4.74 / 26.25 \\
\bottomrule
\end{tabular}
\caption{Median and tail (90th percentile) impact of byzantine solver-suppression attacks aggregated across attack windows.}
\label{tab: byzantine_p90}
\end{table*}

%% file: tables/main-body-results-dos-usdc.tex
\setlength{\tabcolsep}{4pt} % Default value: 6pt

\begin{table*}[ht]
\scriptsize
\rowcolors{2}{gray!10}{white}
\centering
\begin{tabular}{lccccccrrrrr}
\toprule
& & \multicolumn{4}{c}{\textbf{Simulated Values}} & \multicolumn{5}{c}{\textbf{Simulation Results}}\\ 
\cmidrule(rl){3-6} \cmidrule(l){7-11}
Bridge & Src → Dst & S Profit. & Prot. Fee & Max Tx Value & Vol Mul & N. Attacks & N. Fulfillments & Volume Fulfilled (\$) & Net Profit (\$) & $Pr[\text{Profit}]$ \\
\midrule
mayan & solana → ethereum & Real & 0.027 & 10000 & 1 & 514 & 4.36 ± 3.41 & 128.78 ± 117.4 & \cellcolor{red!0}{-25.21 (p90: -7.7)} & 2.1\% \\

mayan & solana → ethereum & 0.018 & Real & 10000 & 1 & 514 & 4.36 ± 3.41 & 128.78 ± 117.4 & \cellcolor{red!0}{-31.03 (p90: -11.44)} & 0.0\% \\

mayan & solana → ethereum & Real & Real & 10000 & 5.61 & 514 & 24.45 ± 19.13 & 722.48 ± 658.6 & \cellcolor{red!0}{-15.59 (p90: 17.36)} & 24.1\% \\

\rowcolor{gray!30}
mayan & solana → ethereum & Real & Real & 10000 & 1 & 514 & 4.36 ± 3.41 & 128.78 ± 117.4 & \cellcolor{red!0}{-28.3 (p90: -8.84)} & 1.8\% \\
\rowcolor{gray!30}
mayan & solana → ethereum & Real & Real & 100000 & 1 & 514 & 4.36 ± 3.41 & 128.78 ± 117.4 & \cellcolor{red!0}{-28.2 (p90: -8.72)} & 1.8\% \\

mayan & solana → ethereum & 1.129 & Real & 10000 & 1 & 514 & 4.36 ± 3.41 & 128.78 ± 117.4 & \cellcolor{red!0}{-29.6 (p90: -10.09)} & 0.0\% \\

mayan & solana → ethereum & Real & 1.3 & 10000 & 1 & 514 & 4.36 ± 3.41 & 128.78 ± 117.4 & \cellcolor{red!9}{-1343.87 (p90: -494.1)} & 0.0\% \\

mayan & solana → ethereum & Real & Real & 10000 & 4.237 & 514 & 18.46 ± 14.45 & 545.66 ± 497.42 & \cellcolor{red!0}{-19.38 (p90: 8.55)} & 17.9\% \\

debridge & solana → ethereum & Real & Real & 10000 & 0.755 & 100 & 5.1 ± 2.38 & 65175.36 ± 116289.88 & \cellcolor{green!46}{144.23 (p90: 275.41)} & \cellcolor{blue!15}96.0\% \\

debridge & solana → ethereum & Real & Real & 10000 & 0.178 & 100 & 1.2 ± 0.56 & 15365.85 ± 27416.69 & \cellcolor{green!6}{20.94 (p90: 42.78)} & \cellcolor{blue!15}67.0\% \\

debridge & solana → ethereum & Real & 0.029 & 10000 & 1 & 100 & 6.75 ± 3.15 & 86324.98 ± 154026.34 & \cellcolor{green!63}{197.22 (p90: 373.53)} & \cellcolor{blue!15}96.0\% \\

debridge & solana → ethereum & 0.381 & Real & 10000 & 1 & 100 & 6.75 ± 3.15 & 86324.98 ± 154026.34 & \cellcolor{green!100}{311.8 (p90: 664.96)} & \cellcolor{blue!15}97.0\% \\

debridge & solana → ethereum & Real & 0.027 & 10000 & 1 & 100 & 6.75 ± 3.15 & 86324.98 ± 154026.34 & \cellcolor{green!63}{198.36 (p90: 375.74)} & \cellcolor{blue!15}96.0\% \\
\rowcolor{gray!30}
debridge & solana → ethereum & Real & Real & 100000 & 1 & 100 & 6.75 ± 3.15 & 86324.98 ± 154026.34 & \cellcolor{green!67}{210.4 (p90: 405.83)} & \cellcolor{blue!15}100.0\% \\

debridge & solana → ethereum & 0.018 & Real & 10000 & 1 & 100 & 6.75 ± 3.15 & 86324.98 ± 154026.34 & \cellcolor{red!0}{-1.56 (p90: 20.8)} & 34.0\% \\
\rowcolor{gray!30}
debridge & solana → ethereum & Real & Real & 50000 & 1 & 100 & 6.75 ± 3.15 & 86324.98 ± 154026.34 & \cellcolor{green!67}{208.99 (p90: 400.77)} & \cellcolor{blue!15}100.0\% \\
\rowcolor{gray!30}
debridge & solana → ethereum & Real & Real & 10000 & 1 & 100 & 6.75 ± 3.15 & 86324.98 ± 154026.34 & \cellcolor{green!63}{196.59 (p90: 374.79)} & \cellcolor{blue!15}96.0\% \\

\bottomrule
\end{tabular}
\caption{Results for $k$ = 1 and AttWindow = 1000s with color-coded net profit considering USDC transfers of less than \$100 in Mayan Swift and USDC transfers of over \$500 in deBridge. Gray rows mark simulations with the real protocol parameters.}
\label{tab: results-usdc}
\end{table*}

%% file: tables/data-model.tex
\begin{table}[ht]
\centering
\scriptsize
\begin{tabular}{@{}ll@{}}
\toprule
\multicolumn{2}{l}{\textbf{Source Transaction (order)}} \\
\midrule
\texttt{blockchain}       & Origin chain where the user deposits funds. \\
\texttt{transaction\_hash} & Unique ID of the transaction on the source chain. \\
\texttt{from}             & User address that initiated the source transaction. \\
\texttt{to}               & Target address of the source transaction. \\
\texttt{fee}              & Fee paid to validators/miners for this transaction. \\
\texttt{value}            & Native value transferred in this transaction. \\
\texttt{timestamp}        & Source transaction timestamp. \\
\addlinespace
\multicolumn{2}{l}{\textbf{Destination Transaction (fulfillment)}} \\
\midrule
\texttt{blockchain}       & Destination chain where funds are released. \\
\texttt{transaction\_hash} & Unique ID of the transaction on the destination chain. \\
\texttt{from}             & Solver address performing the \textit{fulfillment}. \\
\texttt{to}               & Target address of the destination chain. \\
\texttt{fee}              & Fee paid validators/miners for this transaction. \\
\texttt{value}            & Native value transferred in this transaction. \\
\texttt{timestamp}        & Destination transaction timestamp. \\
\addlinespace
\multicolumn{2}{l}{\textbf{Repayment Transaction}} \\
\midrule
\texttt{blockchain}       & Blockchain on which repayment occurs. \\
\texttt{transaction\_hash} & Unique ID of the repayment transaction. \\
\texttt{from}             & Solver address requesting the repayment. \\
\texttt{to}               & Settlement contract address. \\
\texttt{fee}              & Fee paid validators/miners for this transaction. \\
\texttt{value}            & Native value transferred in this transaction. \\
\texttt{timestamp}        & Repayment transaction timestamp. \\
\addlinespace
\multicolumn{2}{l}{\textbf{Intent Information}} \\
\midrule
\texttt{intent\_id}               & Unique intent ID. \\
\texttt{auction\_id}              & Identifier of the auction for this intent. \\
\texttt{depositor}                & User address sending tokens. \\
\texttt{recipient}                & User address receiving tokens. \\
\texttt{src\_contract\_address}   & Address of the original token contract. \\
\texttt{dst\_contract\_address}   & Address of the destination token contract. \\
\texttt{input\_amount}            & Tokens sent. \\
\texttt{input\_amount\_usd}       & Value sent in USD. \\
\texttt{output\_amount}           & Tokens received. \\
\texttt{output\_amount\_usd}      & Value received in USD. \\
\texttt{refund\_amount}           & Refunded tokens. \\
\texttt{refund\_amount\_usd}      & Refunded value in USD. \\
\texttt{refund\_token}            & Token refunded. \\
\texttt{quote\_timestamp}         & Timestamp of the quote generation. \\
\texttt{fill\_deadline}           & Deadline to fill intent. \\
\texttt{exclusivity\_deadline}    & Deadline for exclusive relayer. \\
\texttt{exclusive\_relayer}       & Assigned exclusive solver. \\
\texttt{native\_fix\_fee}         & Fixed fee charged by the protocol in native token. \\
\texttt{native\_fix\_fee\_usd}    & Fixed fee charged by the protocol in USD. \\
\texttt{percent\_fee}             & Percentage fee charged by the protocol. \\
\texttt{percent\_fee\_usd}        & Percentage fee charged by the protocol in USD. \\
\addlinespace
\multicolumn{2}{l}{\textbf{Additional Fields}} \\
\midrule
\texttt{fill\_latency}             & Time taken to fulfill the intent. \\
\texttt{refund\_latency}           & Time taken to refund user. \\
\texttt{auction\_duration}         & Duration of solver auction. \\
\texttt{auction\_start\_latency}   & Timestamp of the start of the auction. \\
\texttt{adjusted\_refund\_fee\_usd} & Refund fee in USD adjusted due to batching. \\
\texttt{adjusted\_dst\_fee\_usd}   & Destination fee in USD adjusted due to batching. \\
\texttt{solver\_profitability}     & Absolute solver profit in USD. \\
\texttt{solver\_profitability\_pct} & Solver profit as percent. \\
\texttt{user\_paid\_protocol\_fees\_usd} & Protocol fees paid by the user in USD. \\
\texttt{user\_paid\_protocol\_fees\_pct} & Protocol fees paid by ther user as percent. \\
\texttt{intent\_value\_usd}        & Intent value in USD. \\
\texttt{src\_symbol}               & Source token symbol. \\
\texttt{dst\_symbol}               & Destination token symbol. \\
\texttt{middle\_src\_symbol}       & Middle source token (if there is a swap). \\
\texttt{middle\_dst\_symbol}       & Middle destination token (if there is a swap). \\
\texttt{middle\_src\_token}       & Middle source token. \\
\texttt{middle\_src\_amount}      & Amount of middle source token. \\
\texttt{middle\_src\_amount\_usd} & Amount of middle source in USD. \\
\texttt{middle\_dst\_token}       & Middle destination token. \\
\texttt{middle\_dst\_amount}      & Amount of middle destination token. \\
\texttt{middle\_dst\_amount\_usd} & Amount of middle destination in USD. \\
\texttt{bridge}                    & Bridge used. \\
\bottomrule
\end{tabular}
\caption{Data model used for an intent-based cross-chain transaction with additional metdadata}
\label{tab:data-model}
\end{table}

\begin{comment}
    -- src blockchain data
src_blockchain,
src_transaction_hash
src_from_address
src_to_address
src_fee
src_value
src_fee_usd
src_timestamp
-- dst blockchain data
dst_blockchain
dst_transaction_hash
dst_from_address
dst_to_address
dst_fee
dst_value
dst_fee_usd
dst_timestamp
-- refund transaction data
refund_blockchain
refund_transaction_hash
refund_from_address
refund_to_address
refund_fee
refund_value
refund_fee_usd
refund_timestamp
-- intent information
intent_id
depositor
recipient
src_contract_address
dst_contract_address
input_amount
input_amount_usd
middle_src_token
middle_src_amount
middle_src_amount_usd
middle_dst_token
middle_dst_amount
middle_dst_amount_usd
output_amount
output_amount_usd
refund_amount
refund_amount_usd
refund_token
quote_timestamp
fill_deadline
exclusivity_deadline
exclusive_relayer
native_fix_fee
native_fix_fee_usd
percent_fee
percent_fee_usd
-- Auction Data
auction_id
auction_first_bid_timestamp
auction_last_bid_timestamp
auction_number_of_bids

-- Additional fields
fill_latency
refund_latency
auction_duration
auction_start_latency
adjusted_refund_fee_usd
adjusted_dst_fee_usd
solver_profitability
solver_profitability_pct
user_paid_protocol_fees_usd
intent_value_usd
user_paid_protocol_fees_pct
src_symbol
dst_symbol
middle_src_symbol
middle_dst_symbol
bridge

\end{comment}

%% file: tables/solver-addresses.tex
\setlength{\tabcolsep}{70pt} % Default value: 6pt

\begin{table}[ht]
\tiny
\centering
\begin{tabular}{l}
\toprule
\addlinespace
\textbf{Mayan} \\ \hline
\texttt{0x6ffc5848c46319e7c6d48f56ca2152b213d4535f} \\
\texttt{0x466b037ace44c0134dcebd965a4a22aed6dea027} \\
\texttt{0xdfd122610a14ac12d934898c02dbec1f72708116} \\
\texttt{0x7c825c6e7e4e1f618ca67e4943cdb41ca00b7f6b} \\
\texttt{0x2977b8919df6a60e93089e0f4231a28899005302} \\
\texttt{0x89c352c56c2caddbb4585731609802fb40867965} \\
\texttt{0x38bf020e39e5a3ef1519c1283f6cac8a6b5851ff} \\
\texttt{0xcbb0cb4492afbcd9963441cc6aea50f35807ff96} \\
\texttt{0x7c825c6E7E4e1F618Ca67E4943CDb41CA00B7f6B} \\
\texttt{0xDfd122610A14Ac12D934898c02dBEc1f72708116} \\
\texttt{0x6Ffc5848c46319e7C6D48f56cA2152B213d4535F} \\
\texttt{0x2977b8919dF6A60E93089E0F4231A28899005302} \\
\texttt{0xcBb0cb4492afbCD9963441cC6AeA50F35807fF96} \\
\texttt{0x466B037ace44C0134Dcebd965A4a22Aed6DEA027} \\
\addlinespace
\textbf{deBridge} \\ \hline
\texttt{0x555ce236c0220695b68341bc48c68d52210cc35b} \\
\texttt{0x1c43ee851156a6ebc643b6a6b13413dcd479fc96} \\
\texttt{0xc4eb49ea01578cb9b1c68ad27f457dbfa0bfbd97} \\
\texttt{0x78b0f42536aeee037deedbb968ffb23cc2c0082e} \\
\texttt{0x1155b614971f16758c92c4890ed338c9e3ede6b7} \\
\texttt{0x7185b9c0c4ffa4eec6ecf100c5bc3583065002ab} \\
\texttt{0x41bc52b02a0f7604cc6fb59ea49e261b60f3ec34} \\
\texttt{0x98ab320ad1f8459d9ea2e83e1fc3ea80504f0eae} \\
\texttt{0x555CE236C0220695b68341bc48C68d52210cC35b} \\
\addlinespace
\textbf{Across} \\ \hline
\texttt{0xeff7337b37c8d217d01cb8223fe497abd75190d5} \\
\texttt{0xcad97616f91872c02ba3553db315db4015cbe850} \\
\texttt{0x394311a6aaa0d8e3411d8b62de4578d41322d1bd} \\
\texttt{0x699ee12a1d97437a4a1e87c71e5d882b3881e2e3} \\
\texttt{0xef1ec136931ab5728b0783fd87d109c9d15d31f1} \\
\texttt{0x41ee28ee05341e7fdddc8d433ba66054cd302ca1} \\
\texttt{0xeeaf25ad4f51fe2f57be2f206c9d8a568a618b99} \\
\texttt{0x84a36d2c3d2078c560ff7b62815138a16671b549} \\
\texttt{0x18105a39db36eb6f865704be858bcc7954c66467} \\
\texttt{0x4d38a9e742450872ed777c4df3cef7d5cbe8e3a8} \\
\texttt{0x15652636f3898f550b257b89926d5566821c32e1} \\
\texttt{0xA36bc6867c9963DE1a5dAaF7882efb5B56899E8c} \\
\texttt{0xa36bc6867c9963de1a5daaf7882efb5b56899e8c} \\
\texttt{0x95b4298680f9418f7e32155ea7069ade25caf5f1} \\
\texttt{0x07ae8551be970cb1cca11dd7a11f47ae82e70e67} \\
\texttt{0x1b59718eafa2bffe5318e07c1c3cb2edde354f9c} \\
\texttt{0x699EE12a1d97437A4A1E87C71e5d882b3881e2e3} \\
\texttt{0xCad97616f91872C02BA3553dB315Db4015cBE850} \\
\texttt{0x428ab2ba90eba0a4be7af34c9ac451ab061ac010} \\
\texttt{0xbe75079fd259a82054caab2ce007cd0c20b177a8} \\
\texttt{0x3d7dc36aa2b542ad239012730dfdb23f03d75be9} \\
\texttt{0x394311A6Aaa0D8E3411D8b62DE4578D41322d1bD} \\
\texttt{0xee51012a186832adc6473ab26b88f5624255a95c} \\
\texttt{0x96ae533814f9a128333a2914a631b9ae690e2b0a} \\
\texttt{0xe7488e7f5c60bfa81d38965c02b4cc5aa6249a11} \\
\texttt{0xefF7337B37c8D217d01cb8223fe497ABD75190d5} \\
\texttt{0x2f9d1d528b055e5fad132f65be128551f1c4a503} \\
\texttt{0x18105A39dB36EB6f865704Be858bcC7954c66467} \\
\texttt{0xeF1eC136931Ab5728B0783FD87D109c9D15D31F1} \\
\texttt{0x27b86c00f28ca06dfdd0b84af577406f323a7935} \\
\texttt{0x4e1bab12e5b9281dbe057f41b67e9a0f505fd37d} \\
\texttt{0x1bfb84669be4018983969f5e369286054805bdd2} \\
\texttt{0xe63bae1ab3eb28eac0cbb5b41f6a09f492654fef} \\
\texttt{0xc7216a9097913f010be78b8d6ee6688b7c30cea4} \\
\texttt{0xaa36fb925d80a4aca267bb1a9d3ef5c1af929cca} \\
\texttt{0x07aE8551Be970cB1cCa11Dd7a11F47Ae82e70E67} \\
\texttt{0xEeAF25aD4f51fE2f57Be2F206C9d8A568A618b99} \\
\texttt{0x15652636f3898F550b257B89926d5566821c32E1} \\
\texttt{0xd2068e04cf586f76eece7ba5beb779d7bb1474a1} \\
\texttt{0xe0bac74ff58ca9d6c40e955e5e4b04129b795b7d} \\
\texttt{0x41ee28EE05341E7fdDdc8d433BA66054Cd302cA1} \\
\texttt{0xbe75079fd259a82054cAAB2CE007cd0c20b177a8} \\
\texttt{0x9e2d2f9953cae6bb53b9805295d7cd6fb47d6ee4} \\
\texttt{0x9a8f92a830a5cb89a3816e3d267cb7791c16b04d} \\
\texttt{0xf136390d409619f18bb96645fb4a2897788c5fd9} \\
\texttt{0xc0ffeebabe5d496b2dde509f9fa189c25cf29671} \\
\texttt{0x9757cb8f382409828ffb1258cb1fe21a43d858ac} \\
\texttt{0x1b59718eaFA2BFFE5318E07c1C3cB2edde354f9C} \\
\texttt{0x89deedeb7b2859dd6ee337f9f692d33b472ce711} \\
\texttt{0x95B4298680F9418F7e32155ea7069ADE25cAf5F1} \\
\texttt{0x3d782c0c69101358a8267ba116c86726fdf35f91} \\
\texttt{0xe59aaf21c4d9cf92d9ed4537f4404ba031f83b23} \\
\texttt{0xc4655836221b1881fbe62908fdcb2cacad41e0c3} \\
\texttt{0xd7d2807b0715aa1af8157574642f4f06c9badda2} \\
\texttt{0x852f57dd17edbb0bedae8c55dd4b20feb3133089} \\
\texttt{0xc31a49d1c4c652af57cefdef248f3c55b801c649} \\
\texttt{0xfe263102682933297cb65dc813e5193249769251} \\
\texttt{0xca73a9a0e16639aa21775de6c81dbe79e6cbc0a3} \\
\texttt{0xee51012A186832adC6473aB26b88f5624255A95C} \\
\texttt{0xd2068e04Cf586f76EEcE7BA5bEB779D7bB1474A1} \\
\bottomrule
\end{tabular}
\caption{All solver addresses identified in our analysis for fulfillments in Ethereum.}
\label{tab: solver_addreses}
\end{table}